\documentclass[aps,pre,amsmath,amssymb,preprint,
superscriptaddress,
]{revtex4-1}
\ifx\pdfoutput\undefined
\usepackage{graphicx}
\else
\usepackage{graphicx}
\usepackage{epstopdf}
\fi
\usepackage[pdftitle={PoF},bookmarks=true,citecolor=red,colorlinks=false]{hyperref}
\usepackage{wasysym}
\usepackage{xcolor}

\usepackage{CJK}
\newcommand{\upd}{\mathrm{\,d}}

\newcommand{\red}[1]{\textcolor{black}{#1}}

\begin{document}

\begin{CJK*}{GB}{gbsn} 
\title{Extremal-point density of scaling processes: Part 
  \uppercase\expandafter{\romannumeral1} \relax from fractal Brownian motion to turbulence in one dimension}

\author{Yongxiang Huang (»ÆÓÀÏé)}
\email{yongxianghuang@gmail.com}
\affiliation{State Key Laboratory of Marine Environmental Science, College of Ocean and Earth Sciences,
Xiamen University, Xiamen 361102,  China}

\author{Lipo Wang (ÍõÀûÆÂ)}
\affiliation{UM-SJTU Joint Institute, Shanghai JiaoTong University, Shanghai, 200240, China}

\author{F. G. Schmitt}%
\affiliation{CNRS, Univ. Lille, Univ. Littoral Cote d'Opale, UMR 8187, LOG, Laboratoire d'Oc\'eanologie et de G\'eosciences, F 62 930 Wimereux, France}

\author{Xiaobo Zheng  (֣С²¨)}
\affiliation{Department of Mechanics, Tianjin University, 300072 Tianjin, China}
\author{Nan Jiang (½ªéª)}
\affiliation{Department of Mechanics, Tianjin University, 300072 Tianjin, China}

\author{Yulu Liu (ÁõÓî½)}%
\affiliation{Shanghai Institute of Applied Mathematics and Mechanics, Shanghai University,
Shanghai 200072,  China}

\date{\today}

\begin{abstract}
In recent years several local extrema based methodologies have been proposed to investigate either the nonlinear or the nonstationary time series for scaling analysis. In the present work we study systematically the distribution of the local extrema for both  synthesized scaling processes and  turbulent velocity data from experiments. The results show that for the fractional Brownian motion (fBm) without intermittency correction the measured extremal point density (EPD) agrees well with a theoretical prediction. For a multifractal random walk (MRW) with the lognormal statistics, the measured EPD is independent with the intermittency parameter $\mu$, suggesting that the intermittency correction does not change the distribution of extremal points, but change the amplitude. By introducing a coarse-grained operator, the power-law behavior of these scaling processes is then revealed via the measured EPD for different scales. For fBm the scaling exponent $\xi(H)$ is found to be $\xi(H)=H$, where $H$ is Hurst number, while for MRW $\xi(H)$ shows a linear relation with the intermittency parameter $\mu$. Such EPD approach is further applied to the turbulent velocity data obtained from a wind tunnel flow experiment with the Taylor scale $\lambda$ based Reynolds number $Re_{\lambda}= 720$, and a turbulent boundary layer with the momentum thickness $\theta$ based Reynolds number $Re_{\theta}= 810$. A scaling exponent $\xi\simeq 0.37$ is retrieved for the former case. For the latter one, the measured EPD shows clearly four regimes, which agree well with the four regimes of the turbulent boundary layer structures.
\end{abstract}

\maketitle
\end{CJK*}

\section{Introduction}
Multiscale statistics is recognized as one of the most import features of complex dynamical systems. Several methodologies have been put forward to characterize the multiscale property, such as structure function analysis proposed by \citet{Kolmogorov1941}, wavelet-based approaches \cite{Muzy1993PRE,Lashermes2005wavelet},  the Hilbert-Huang transform \cite{Huang1998EMD,Huang2008EPL} and multi-level segment analysis \cite{Wang2015JSTAT}. The local extremal point (see definition below) plays important roles in multiscale characterization \citep{Huang1998EMD,Huang2008EPL,Wang2015JSTAT,Muzy1993PRE}. For example, in the Hilbert-Huang transform, local extrema are used to construct the upper/lower envelope \citep{Huang1998EMD}; in multi-level segment analysis, a structure function is defined conditionally on the segments between consecutive extremal points \citep{Wang2015JSTAT}. Experimental results suggest that these two methods can overcome some potential shortcomings of the conventional structure function \cite{Wang2015JSTAT,Schmitt2016Book,Huang2010PRE}, such as scale mixing. The scale and the corresponding scaling or multifractality nature are embedded in the local extremal point statistics, which definitely deserves further studies.

\red{The distribution of the local extremal point is associated with the dynamical behavior of the considered processes. Concerning a discrete time series $x(t_i)$, $i=1,2,3\cdots N$, and a sampling frequency $f_s$,  the local extremal point (either local maxima or minima) satisfies the following relation
\begin{equation}
x_t(t_{i+1})x_t(t_{i})< 0 , \label{eq:extrema}
\end{equation} 
where $x_t(t_i)=(x(t_{i+1})-x(t_i))/(t_{i+1}-t_i)$ is the local slope of $x(t_i)$. 
This property  is used for direct counting of the number of extrema.
 Clearly the local extrema correspond to the zero-crossing points of the
  first-order derivative of $x(t_i)$. If $x(t_i)$ acts as the turbulent velocity, $x_t(t_i)$ is then the acceleration. The local extreme is thus an indicator of the sign change of the acceleration/forcing, showing the dynamical property of $x(t_i)$ \citep{Wang2015JSTAT}. 
 Theoretically, \citet{Rice1944BSTJ} proved  that for a stationary continuous  process $x(t)$, if $x(t)$ and $x_t(t)$ are statistically independent and  Gaussian distributed, the zero-crossing ratio (ZCR) per second of $x(t)$ denoted as $N_0$ can  be expressed as
\begin{equation}
\red{N_0=\frac{1}{\pi}\left(\frac{\langle x_t^2(t) \rangle_t}{\langle x^2(t)\rangle_t}\right)^{1/2}}, \label{eq:Rice}
\end{equation}
where $\langle \,\rangle_t$ denotes the sample average with respect to $t$. 
Similarly the   ZCR of the first order derivative, i.e., $x_{t}(t)$, denoted as $N_1$ can be written as \citep{Longuet1958}
\begin{equation}
N_1=\frac{1}{\pi}\left(\frac{\langle x_{tt}^2(t) \rangle_t}{\langle x_t^2(t)\rangle_t}\right)^{1/2},
\end{equation}
where $x_{tt}(t)$ is the second-order derivative. The corresponding extremal-point-density (EPD), e.g., the ratio between number of extremal points and the total data length, is then written as
\begin{equation}
\mathcal{I}=\frac{N_1}{f_s}=\frac{1}{\pi f_s}\left(\frac{\langle x_{tt}^2(t) \rangle_t}{\langle x_t^2(t)\rangle_t}\right)^{1/2},\label{eq:RiceMM}
\end{equation}
where $f_s$ is the sampling frequency of the discrete process. Later \citet{Ylvisaker1965} showed that in Eq.\,\eqref{eq:Rice} for any continuous stationary Gaussian process with finite $N_0$ the statistical independence between $x(t_i)$ and $x_t(t_i)$ need not to be invoked.
More detail about the ZRC and the Rice formula can be found in Ref.\,\citep{Leadbetter2012book}}

\citet{Toroczkai2000PRE} proposed \red{another} theory to estimate the
EPD as follows. Assume $p(\phi_1,\phi_2)$, the joint probability density function (pdf) of the distribution of two neighbor slopes $\phi_1$ and $\phi_2$, where $\phi_i=x(t_{i+1})-x(t_i)$, satisfies a joint Gaussian distribution as
\begin{equation}
p(\phi_1,\phi_2)=\frac{1}{2\pi\sqrt{D}}\exp\left[ -\frac{d}{2D} \left( \phi_1^2 +\phi_2^2-2\frac{c}{d}\phi_1\phi_2\right)\right], \label{eq:jpdf}
\end{equation}
where $\langle \phi_1^2\rangle=\langle \phi_2^2\rangle=d>0$, $\langle\phi_1\phi_2\rangle=c$ and $D=d^2-c^2>0$. Then the process EPD $\mathcal{I}=\frac{N_e}{N}$, where $N_e$ and $N$ are respectively the number of extreme points and the data length, is determined by
\begin{equation}
\mathcal{I}=\frac{1}{\pi}\arccos\left(\frac{\langle \phi_1\phi_2 \rangle}{\langle\phi_1^2 \rangle}\right)=\frac{1}{\pi}\arccos\left(\frac{c}{d}\right). \label{eq:TEPD}
\end{equation}
More detailed derivation of this formula are found in Ref. \citep{Toroczkai2000PRE}.

Specifically for the turbulent velocity signal, \citet{Liepmann1949} pointed out theoretically that the ZRC could be related with the Taylor microscale $\lambda$ via the following relation
\begin{equation}
\red{\lambda
=\frac{1}{\pi N_0}},
\end{equation}
which has been verified experimentally  in 
Ref.\,\cite{Sreenivasan1983JFM,Kailasnath1993PoF}. In the turbulence literature, the local extrema or zero-crossing or level-crossing are also related with the dissipation scale or intermittency \cite{Ho1997JFM,Poggi2009PoF,Poggi2010BLM,Cava2012JGR,Yang2016CCP}. For example, \citet{Ho1997JFM} proposed a peak-valley-counting technique to detect the dissipation scale and found that the most probable scale equals the wavelength at the peak of the dissipation spectrum. \citet{Yang2016CCP} studied the local zero-crossings and their relation with inertial range intermittency for the transverse velocity and passive scalar in an incompressible isotropic turbulent field. They demonstrated that the most intermittent regions for the transverse velocity are inclined to be vortex dominated.

In this paper, the statistics of EPD of several representative scaling processes will be investigated. We first verify Eqs. \eqref{eq:RiceMM} and \eqref{eq:TEPD} using synthesized fractional Brownian motion (fBm) and multifractal random walk (MRW) in Sec.\,\ref{sec:NV}. A coarse-grained algorithm is then proposed to detect the respective scaling behavior. In Sec.\,\ref{sec:Applications}, the real data obtained from various typical turbulent flows are analyzed and the main conclusions are summarized in Sec.\,\ref{sec:Conclusion}.

\section{Numerical Validation}\label{sec:NV}

\subsection{Fractional Brownian motion}

\begin{figure}[!htb]
\centering
 \includegraphics[width=0.85\linewidth]{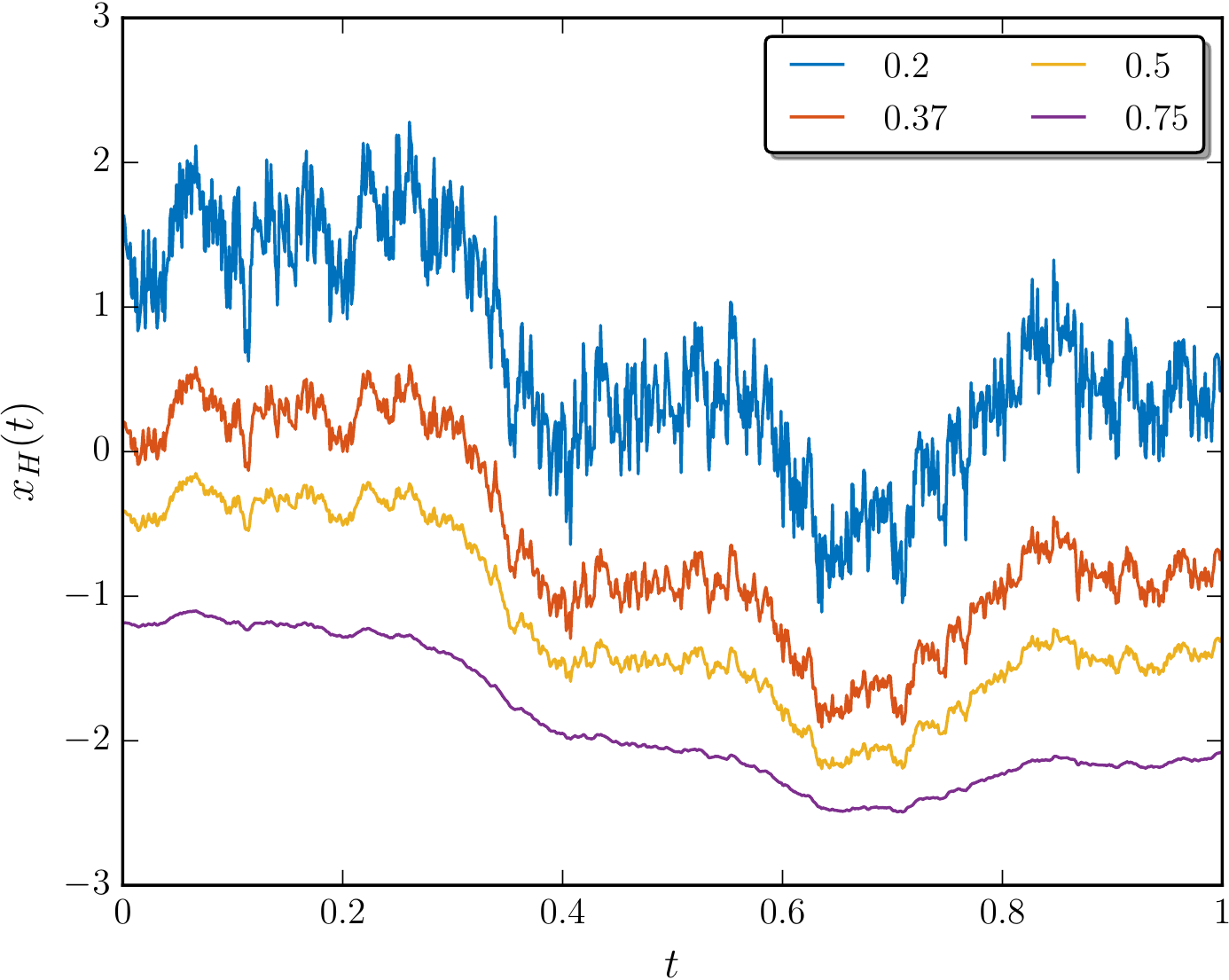}
 \caption{(Color online) Synthesized fractional Brownian motion data for various Hurst number $H$ using the same random numbers in the Wood-Chan algorithm with $1024$ data points in each realization. Visually, with the increasing of $H$, the fBm curve becomes more and more smooth, i.e. less extremal points. For display clarity, these curves have been vertically shifted.}\label{fig:fBm}
\end{figure}

\begin{figure}[!htb]
\centering
 \includegraphics[width=0.85\linewidth]{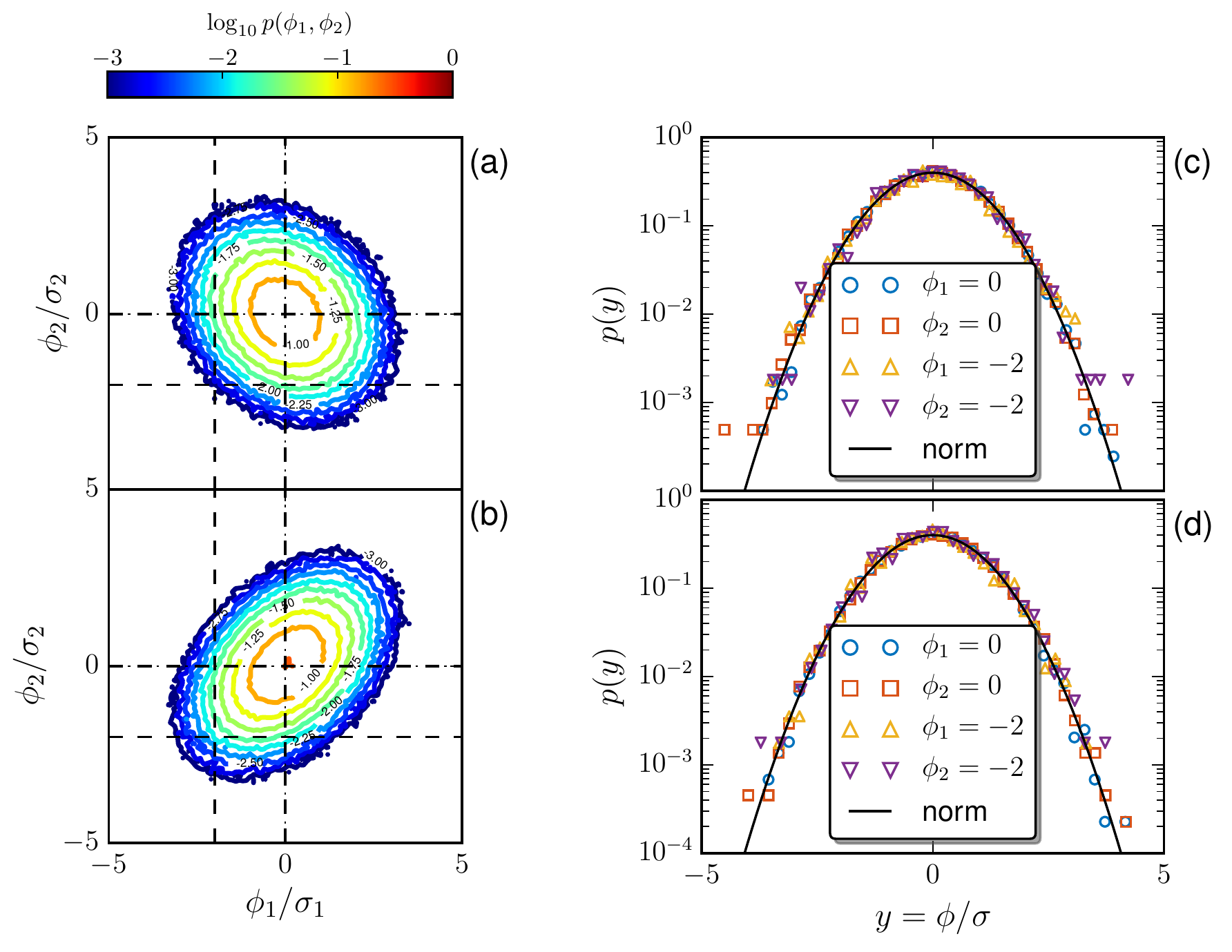}
  \caption{(Color online) Joint pdf $p(\phi_1,\phi_2)$ for a) $H=1/3$ and b) $H=0.75$. The corresponding normalized conditional pdf $p(y)$ at various $\phi_1$ and $\phi_2$ for c) $H=1/3$ and d) $H=0.75$. For comparison, the normal distribution is illustrated as a solid line. Dashed lines in a) and b) denote the calculated $\phi_1$ and $\phi_2$ value.}\label{fig:fbmjpdf}
\end{figure}

\begin{figure}[!htb]
\centering
 \includegraphics[width=0.85\linewidth]{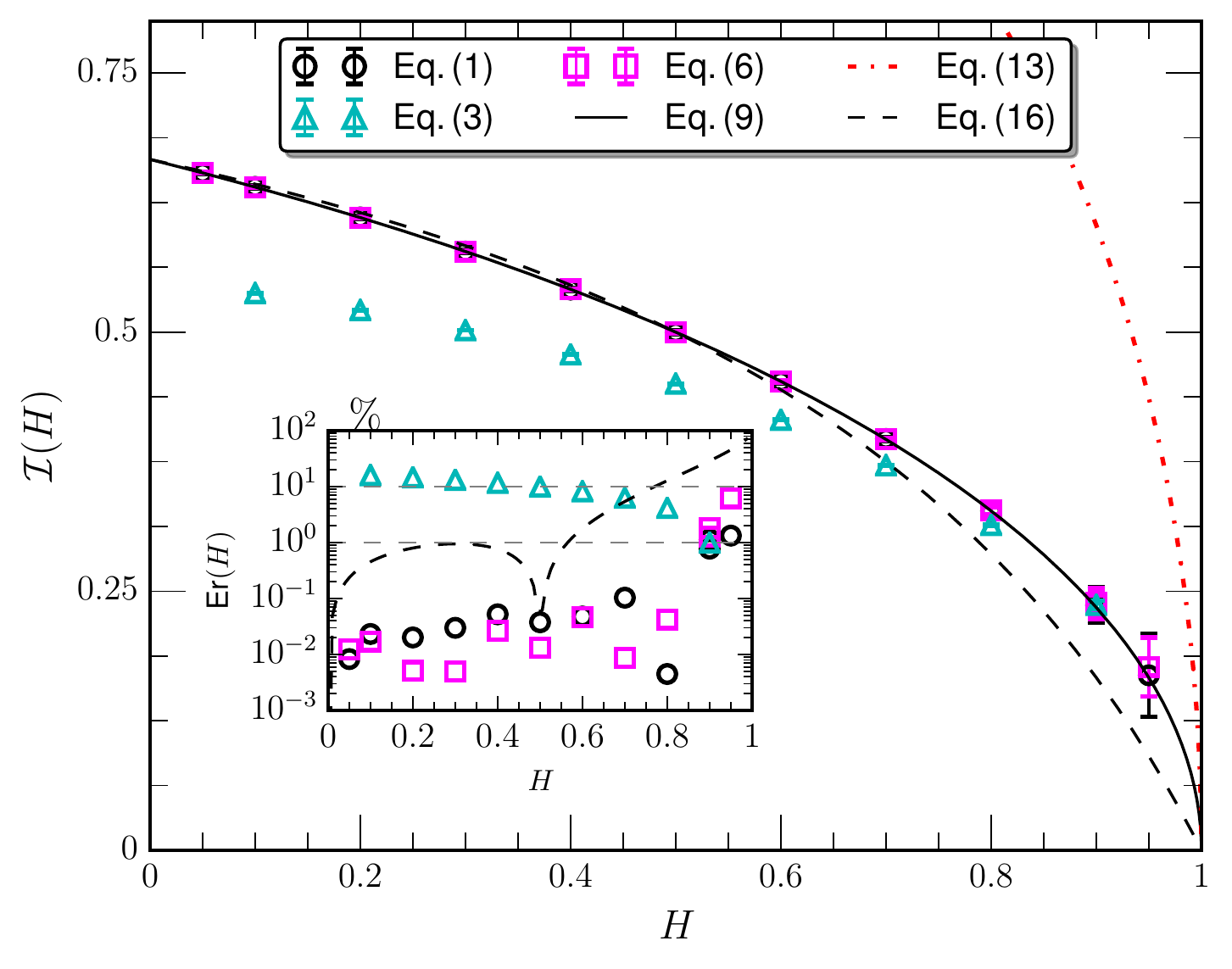}
  \caption{(Color online) \red{ Comparison the measured 
  $\mathcal{I}(H)$  from direct counting, Eq.\,\eqref{eq:extrema} ($\ocircle$),  Eq.\,\eqref{eq:TEPD} ($\square$) via calculating $\phi_i$ and Eq.\,\eqref{eq:RiceMM} via calculating $\langle x_t(t_i)^2\rangle$ and $\langle x_{tt}(t_i)^2\rangle$ ($\triangle$), with theoretical predictions by Eqs.\,\eqref{eq:MF1} (solid line: $\frac{1}{\pi}\arccos(2^{2H-1}-1)$), \eqref{eq:MF4} (dashed-dotted line: $2\sqrt{\frac{2-2H}{4-2H}}$), and \eqref{eq:MF3} (dashed line: $\frac{2-2H}{3-2H}$), respectively. The inset shows the relative error between Eq.\,\eqref{eq:MF1} and others.}
   }\label{fig:numberratio}
\end{figure}

\subsubsection{Extremal-point-density of fractional Brownian motion}
First fBm is considered here as a toy model for a better understanding of the Hurst-dependence of EPD. FBm is a generalization of the classical Brownian motion.
It was introduced by \citet{Kolmogorov1940} and extensively studied by Mandelbrot and co-workers in the 1960s \cite{Mandelbrot1968}. Since then, fBm became to be a classical mono-scaling stochastic process in many fields \cite{Beran1994,Rogers1997,Doukhan2003}. Its first-order derivative is the so-called fractional Gaussian noise (fGn) with the covariance as
\begin{equation}
\rho_H(\tau)=\red{\langle x'(t)x'(t+\tau) \rangle_t}=\frac{\sigma^2}{2}\left( \vert \tau-1\vert^{2H}-2 \vert \tau\vert^{2H}+\vert \tau+1\vert^{2H} \right), \label{eq:fGnACF}
\end{equation}
where $\tau$ is the separation lag, $\rho_H(0)=\sigma^2$ is the variance and $H$ is the Hurst number. Accordingly it yields $\langle \phi_1 \phi_2\rangle =\rho_H(1)=\sigma^2(2^{2H-1}-1)$, and $\langle \phi_1^2\rangle=\rho_H(0) =\sigma^2$. Thus from Eq.~\eqref{eq:TEPD} the process EPD can be written as
\begin{equation}
\mathcal{I}(H)=\frac{1}{\pi}\arccos\left(2^{2H-1}-1\right). \label{eq:MF1}
\end{equation}
\red{Clearly, Eq.\,\eqref{eq:MF1} satisfies the requirement $\mathcal{I}(1/2)=1/2$, and $\mathcal{I}(H)\vert_{\lim_{H\rightarrow1}}=0$; meanwhile, it predicts $\mathcal{I}(H)\vert_{\lim_{H\rightarrow0}}=2/3$.}

\red{As aforementioned, the EPD of $x(t)$ can be associated with the ZCR of $x'(t)$. We introduce here a $n$th-order spectral moment $\omega(n)$ for the fGn, which is written as,
\begin{equation}
\omega(n)=\int_0^{f_s}E_H(f)f^n \upd f \label{eq:omega}
\end{equation} 
where $f_s$ is the sampling frequency of the considered  discrete time series. The EPD can be related  with $E_H(f)$ via the following exact relation \citep{Rice1944BSTJ}, i.e.,
\begin{equation}
\mathcal{I}(H)=\frac{2}{f_s}\times\sqrt{\frac{\omega(2)}{\omega(0)}}\label{eq:RiceM}
\end{equation}
Because for fGn,
\begin{equation}
E_H(f)\propto f^{1-2H} , \label{eq:FfGn}
\end{equation}
Substituting Eq.~\eqref{eq:FfGn} into \eqref{eq:RiceM} then yields
\begin{equation}
\mathcal{I}(H)=2\times\sqrt{\frac{2-2H}{4-2H}},\label{eq:MF4}
\end{equation}
It satisfies  $\mathcal{I}(H)\vert_{\lim_{H\rightarrow1}}=0$; meanwhile predicts $\mathcal{I}(H)\vert_{\lim_{H\rightarrow0}}=\sqrt{2}$, and $\mathcal{I}(1/2)=\sqrt{4/3}$, which does not consist with the requirement $\mathcal{I}\le 1$.
}

\red{An  energy weighted  mean frequency can be defined as \citep{Huang1998EMD,Huang2009Hydrol} 
 \begin{equation}
\tilde{f}(H)=\frac{\omega(1)}{\omega(0)}. \label{eq:EWMF}
 \end{equation}
 Substituting Eq.~\eqref{eq:FfGn} into \eqref{eq:EWMF} then yields,
\begin{equation}
\tilde{f}(H)=f_s\times \frac{2-2H}{3-2H}, \label{eq:MF2}
\end{equation}
Phenomenologically, we assume here  that the EPD of fGn can be related with the mean frequency $\tilde{f}(H)$ as, 
\begin{equation}
\mathcal{I}(H)=\frac{\tilde{f}(H)}{f_s}=\frac{2-2H}{3-2H}. \label{eq:MF3}
\end{equation}
Note that both equations \eqref{eq:MF1} and \eqref{eq:MF3} satisfy the requirements $\mathcal{I}(1/2)=1/2$ and $\mathcal{I}(H)\vert_{\lim H\rightarrow 1}=0$; meanwhile another limit case $\mathcal{I}(H)\vert_{\lim H\rightarrow 0}=2/3$ can be predicted, while Eq.\,\eqref{eq:MF4} only satisfies the limit case $H\rightarrow 1$. 
}

Numerically a Fourier-based Wood-Chan algorithm \citep{Wood1994} was used to generate the fBm data in the range $0\le H\le1$ for 100 realizations, each of which having the data length of $L$. Figure \ref{fig:fBm} shows the synthesized fBm data for various $H$ with $L=1024$ data points. For display clarity, the fBm curves have been vertically shifted. It need to mention that here for each realization the different $H$ cases used the same random numbers in the algorithm for a detailed comparison. Visually, the larger $H$ is, the smoother the process; or in other words, the EPD $\mathcal{I}(H)$ decreases with $H$.

Figure \ref{fig:fbmjpdf} shows the contour line of measured joint pdf $p(\phi_1,\phi_2)$ for a) Hurst number $H=1/3$, and b) $H=0.75$, respectively, with data length $L=100,000$ data points. The inclined ellipse contour lines are centered at $[0,0]$, as indicated by Eq.\,\eqref{eq:jpdf}. Several conditional pdfs $p(\phi_1\vert\phi_2)$ (or $p(\phi_2\vert\phi_1)$) at various $\phi_1$ and $\phi_2$ are shown in Fig.\,\ref{fig:fbmjpdf}\,c) and d), where the solid line represents the normal distribution. Clearly they are in good agreement, showing the applicability of \citet{Toroczkai2000PRE}'s theory.

Figure \ref{fig:numberratio} shows $\mathcal{I}(H)$ obtained from direct counting (Eq.\,\eqref{eq:extrema}, $\ocircle$), \red{Eq.\,\eqref{eq:RiceMM} by calculating $\langle x_t(t_i)^2\rangle $ and $\langle x_{tt}(t_i)^2\rangle $ ($\triangle$),    Eq.\,\eqref{eq:TEPD} by estimating $\phi_i$  ($\square$), and theoretical predictions by  Eq.\,\eqref{eq:MF1} (solid line), Eq.\,\eqref{eq:MF4} (dashed dotted line) and Eq.\,\eqref{eq:MF3} (dashed line), respectively. The error bar is the standard deviation from 100 realizations. The direct measured $\mathcal{I}(H)$ ($\ocircle$) agrees well with Eq.\,\eqref{eq:MF1}, but deviates from Eq.\,\eqref{eq:MF3} when $H\ge 0.6$. Note that both the estimator by Eq.\,\eqref{eq:RiceMM} and theoretical Eq.\,\eqref{eq:MF4} are far from the direct measurement.  The discrepancy of the Rice's formula and the measurement might be due to the fact that the fBm process is not differentiable. }

Since $\mathcal{I}_{\mathrm{T}}(H)$ provided by Eq.\,\eqref{eq:MF1} agrees very well with the direct counting results from Eq.\,\eqref{eq:extrema}, to characterize the measurement error, we introduce here the following relative error by taking Eq.\,\eqref{eq:MF1} as the reference case
\begin{equation}
\mathrm{Er}(H)=\frac{\vert \mathcal{I}(H)-\mathcal{I}_T(H)\vert}{\mathcal{I}_T(H)}\times 100\%,
\end{equation}
which is presented in the inset in  Fig.\,\ref{fig:numberratio}, where $1\%$ and $10\%$ are illustrated by a dashed line. For most of  the values of $H$, $\mathrm{Er}(H)$ is less than $1\%$. Eq.\,\eqref{eq:MF3} has a $\le 1\%$ relative error when $H\le 0.6$, and a $\le 10\%$ error when $H\le 0.8$. Probably such deviation could result from the violation of the convergency assumption that is used to obtain Eq.\,\eqref{eq:omega}.

\subsubsection{Finite length effect}

\begin{figure}[!htb]
\centering
 \includegraphics[width=0.85\linewidth]{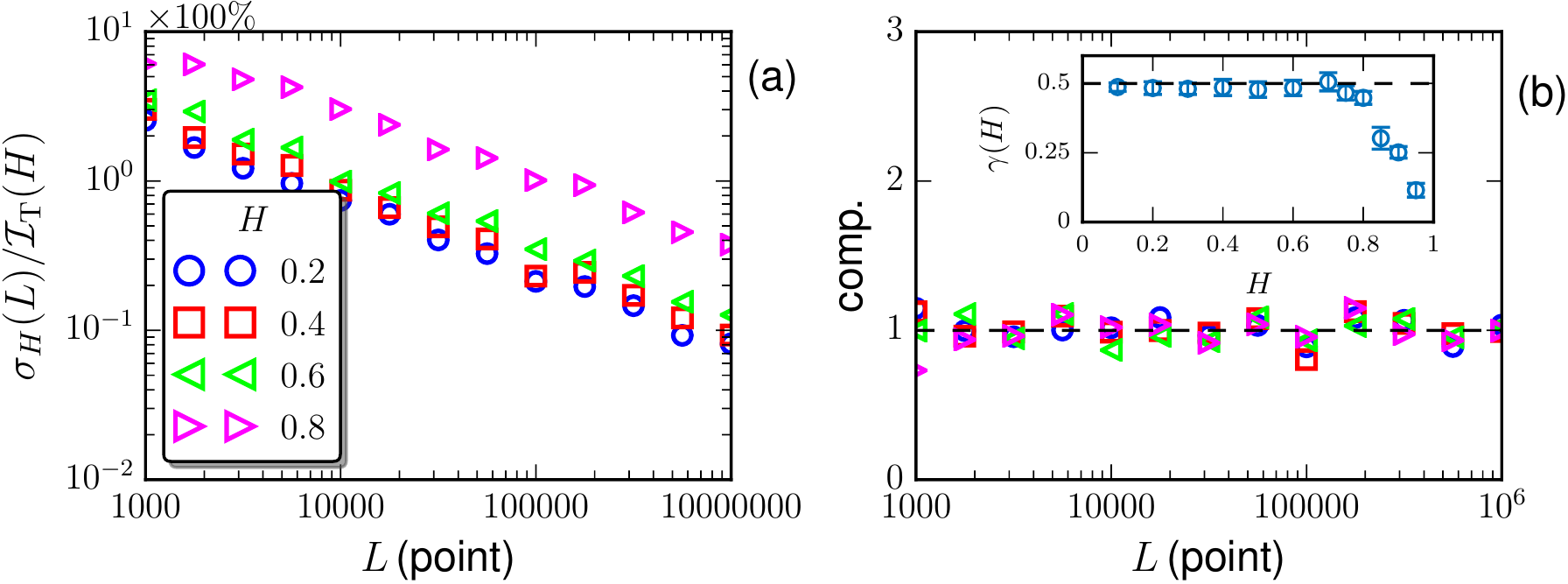}
  \caption{(Color online) a) Relative standard deviation $\sigma_H(L)/\mathcal{I}_{\mathrm{T}}(H)\times 100\%$  for various $H$. b) The compensated data using fitted parameters to emphasize the power-law behavior. The inset shows the measured scaling exponent $\gamma(H)$, where the dashed line showing the value $1/2$, expected from the central limit theorem. The errorbar is the $95\%$ fitting confidence interval.}\label{fig:fBmFLE}
\end{figure}

To consider the influences of the finite length $L$, the synthesized fBm data were generated with different data length $L$ in the range $10^3\sim 10^6$, \red{where only the direct counting method to estimate EPD is considered.}
Figure \ref{fig:fBmFLE}\,a) shows the measured relative standard deviation $\sigma_H(L)/\mathcal{I}_{\mathrm{T}}
(H)\times100\%$  from 100 realizations, where $\mathcal{I}_{\mathrm{T}}(H)$ is the EPD provided by Eq.\,\eqref{eq:MF1}. A power-law decay is observed for all $H$ as
\begin{equation}
\sigma_H(L)\propto L^{-\gamma(H)}.
\end{equation}
To emphasize the observed power-law behavior, corresponding compensated curves using data fitting are shown in 
Fig.\,\ref{fig:fBmFLE}\,b), where the inset shows the measured scaling exponent
$\gamma(H)$.  A clear plateau
confirms that the measured $\mathcal{I}(H)$ converges to the theoretical
value with a power-law rate. It is interesting to note that when $H\le0.7$ $\gamma(H)$ is in a good agreement with the theoretical prediction, $1/2$, which can be expected from the central limit theorem. Typically for $H\le 0.7$, the relative error is below $1\%$ when $L\ge 10,000$ data point, which is easy to be satisfied by the real data set.

\subsubsection{Coarse-grained effect}

\begin{figure}[!htb]
\centering
\includegraphics[width=0.85\linewidth]{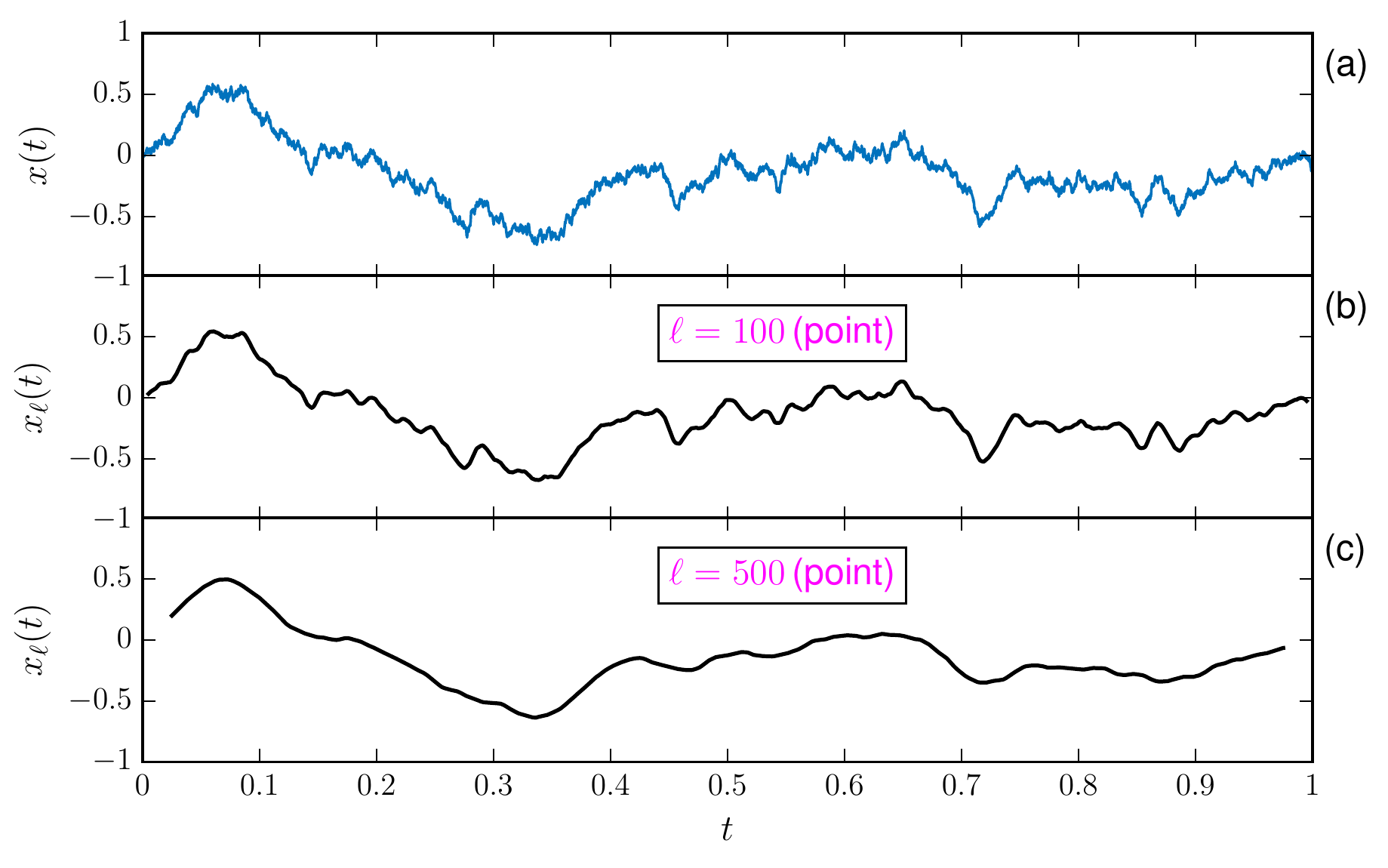}
  \caption{(Color online) Illustration of the filtering effect: a) raw data with $H=0.5$ and $10,000$ data points; b) $\ell=100$ data points; c) $\ell=500$ data points. Increasing $\ell$ removes more local extremal points.}\label{fig:Filtering}
\end{figure}

\begin{figure}[!htb]
\centering
 \includegraphics[width=0.85\linewidth]{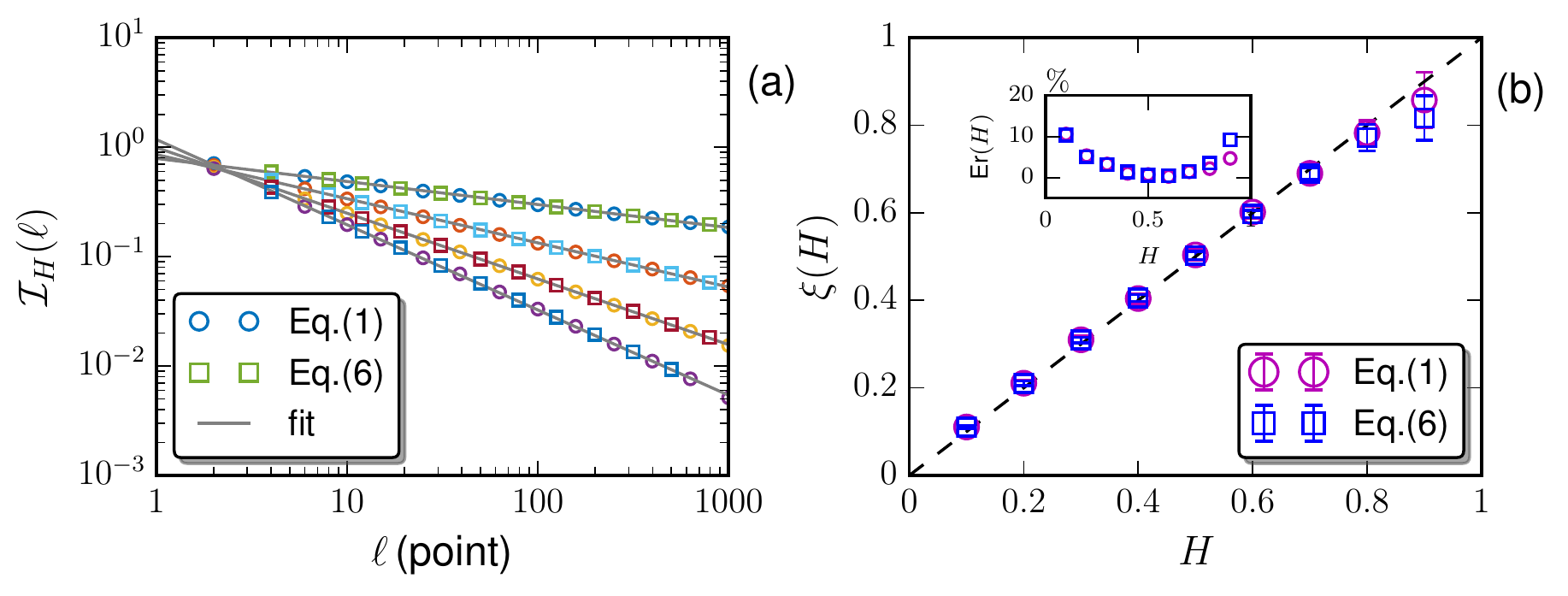}
  \caption{(Color online) a) Measured  $\mathcal{I}(H,\ell)$ with $H=0.2,0.4,0.6$ and $0.8$. The solid line is a power-law fit. For display convenience, $\mathcal{I}(H,\ell)$ is normalized by $\mathcal{I}(H,0)$.
  b) The measured scaling exponent $\xi(H)$ versus $H$. The errorbar is the standard deviation obtained from 100 realizations. The inset shows the relative error $\mathrm{Er}(H)$ versus $H$. }\label{fig:FE}
\end{figure}

For the data from the real world, extremal points can be largely contaminated by noises from different sources. The low-pass filter technique is a commonly adopted remedy for such problem. In the turbulence community the similar coarse-grained idea plays an important role in the multifractal analysis, for instance when considering the energy dissipation rate along the Lagrangian trajectory \cite{Huang2014JFM}. For a continuous process, e.g. $x(t)$, the coarse-grained variable is defined as
\begin{equation}
x_{\ell}(t)=\frac{1}{M}\int_{0\le t\le \ell}  x(t+t')G(t')\upd t' ,\label{eq:CG}
\end{equation}
in which $\ell$ is the coarse-grained scale and $G(t')$ is the filtering kernel, and $M=\int_{0\le t'\ell}G(t')\upd t'$. A simple choice is the hat function as
\begin{equation}
G(t')= \left\{
\begin{array}{lll}
& 1,\quad &\textrm{ }0\le t'\le \ell\\
&0,\quad & \textrm{others }.
\end{array}
\right.
\end{equation}
A discrete version of Eq.\,\eqref{eq:CG} is written as, 
\begin{equation}
x_{\ell}(t_i)=\frac{1}{M}\sum_{j=0}^{\ell-1}x(t_i+j)G(j)
\end{equation}
where $M=\sum_{j=0}^{\ell-1}G(j)$, and $\ell$ is the coarse-grained scale in data point.


Figure \ref{fig:Filtering} illustrates an example of the filtering effect for $H=1/2$ with $10,000$ data points, where the number of local extrema decreases with the coarse-grained scale $\ell$. For various $H$ and $\ell$,  $\mathcal{I}(H,\ell)$ after the coarse-grained operation were calculated with 100 realizations. Figure \ref{fig:FE}\,a) shows $\mathcal{I}(H,\ell)$ with $1\le \ell\le 1,000$ from direct counting ($\ocircle$) and Eq.\,\eqref{eq:TEPD}, where the dependence of the correlation coefficients in Eq.~\eqref{eq:TEPD} on $\ell$ can be obtained either numerically or theoretically.

For display convenience, $\mathcal{I}(H,\ell)$ has been normalized by $\mathcal{I}(H,0)$. Visually, Eq.\,\eqref{eq:TEPD} provides the same value as direct counting since the joint pdf $p(\phi_1,\phi_2)$ can be well described by the joint Gaussian distribution, i.e. Eq.\,\eqref{eq:jpdf}. The following power-law behavior is observed
\begin{equation}
\mathcal{I}(H,\ell)\propto \ell^{-\xi(H)}. \label{eq:PL}
\end{equation}
Fig.\,\ref{fig:FE}\,b) suggests that the scaling exponent $\xi(H)=H$, with the errorbars as the standard deviation from 100 realizations. The inset shows the relative error $\mathrm{Er}(H)$ between $\xi(H)$ and $H$. Clearly  $\mathrm{Er}(H)$ is less than $10\%$, implying a rather good estimation of the Hurst number $H$. For instance, for the turbulent velocity case of $H=1/3$, $\mathrm{Er}(1/3)\simeq 2\%$.

\subsection{Multifractal random walk with lognormal statistics}

\begin{figure}[!htb]
\centering
 \includegraphics[width=0.85\linewidth]{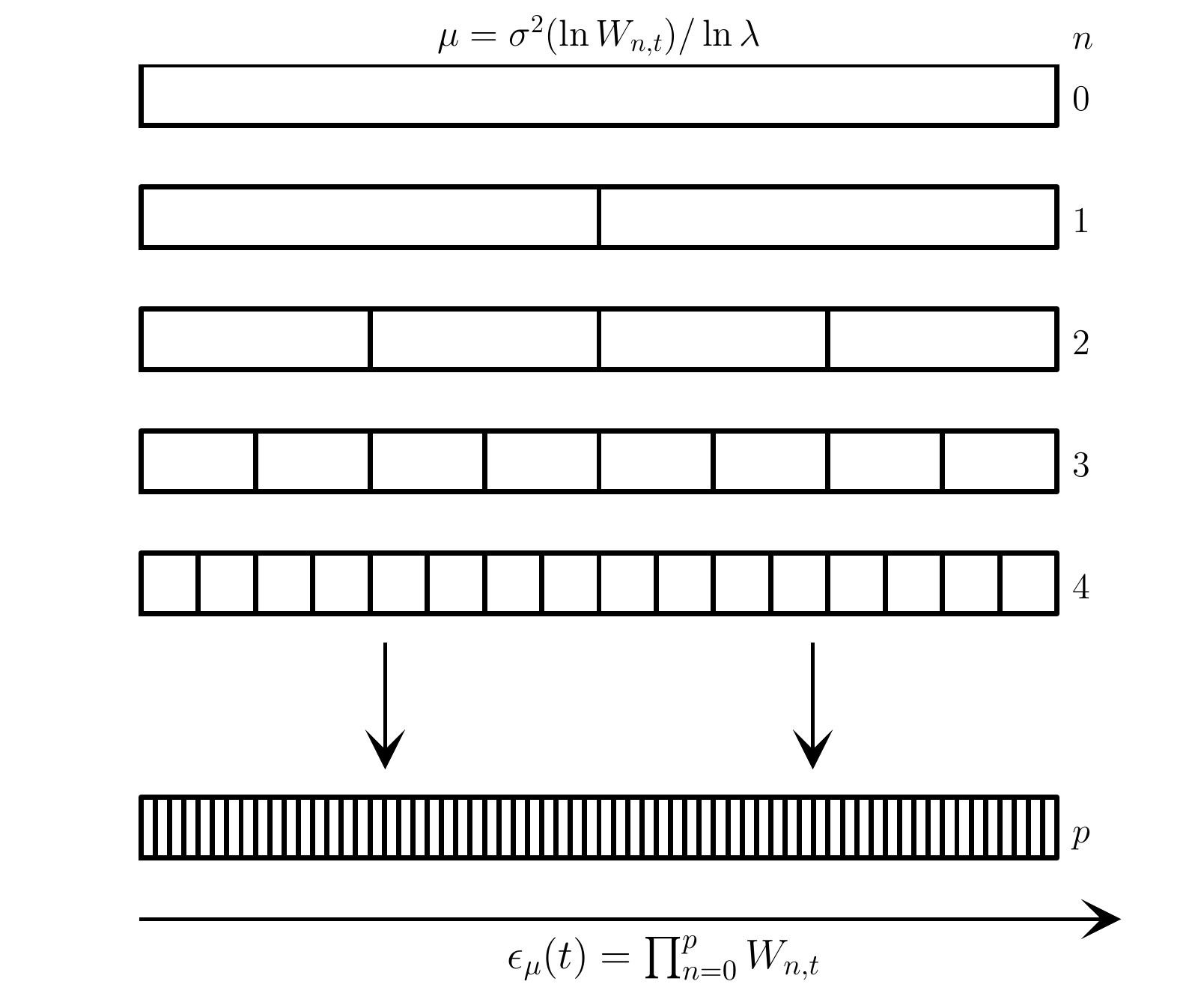}
  \caption{(Color online) Illustration of the discrete MRW cascade process. Each step is associated with a scale ratio of $2$. After $p$ steps, the total scale ratio is $2^p$. The synthesized multifractal measure $\epsilon_{\mu}(t)$ has a lognormal statistics. The corresponding intermittency correction is controlled by the parameter \red{$\mu=\sigma^2(\ln W_n)/\ln \lambda$}.}\label{fig:MFmeasure}
\end{figure}

\begin{figure}[!htb]
\centering
 \includegraphics[width=0.85\linewidth]{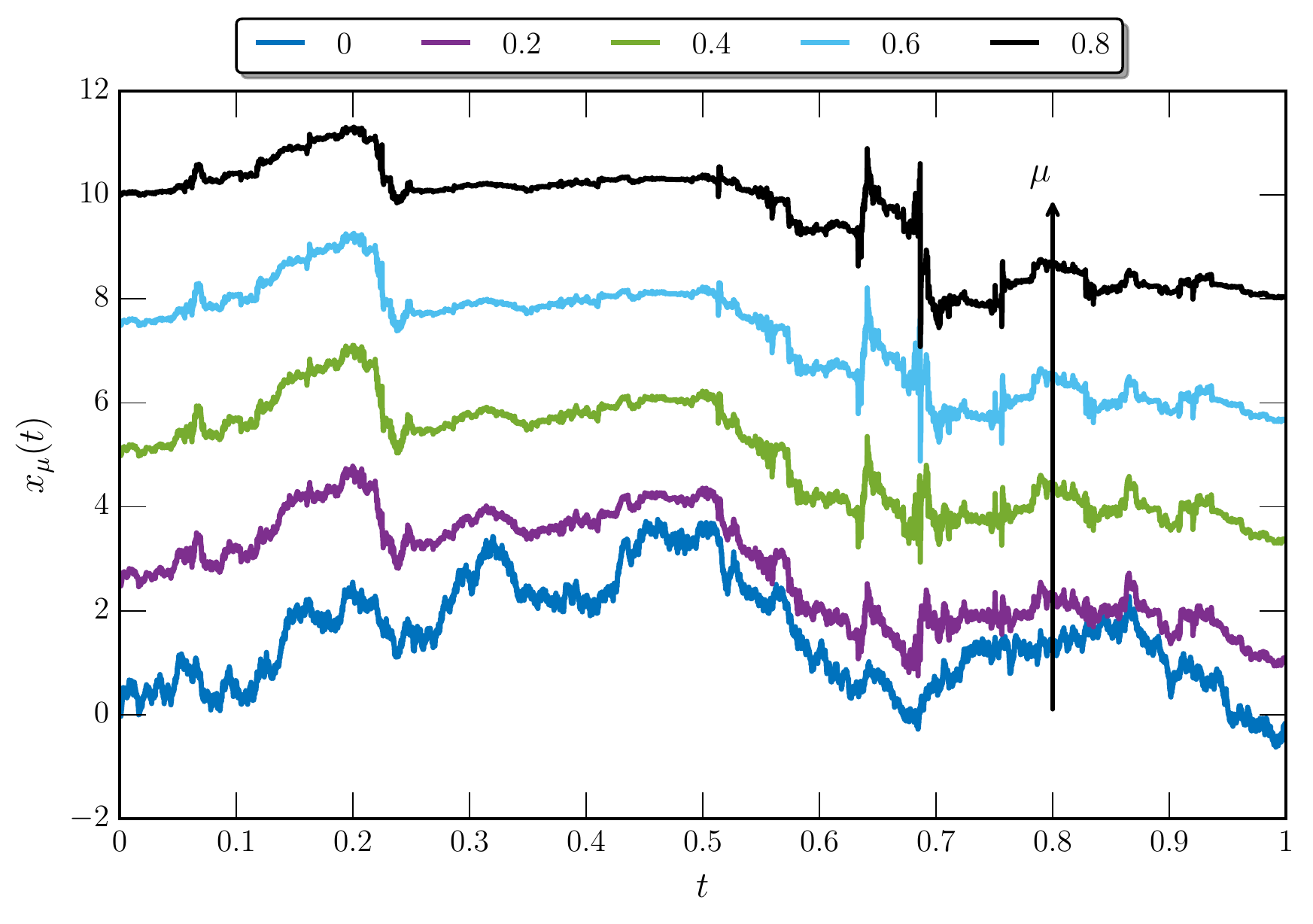}
  \caption{(Color online) Illustration of the synthesized MRW $x_{\mu}(t)$, for one realization with various intermittency parameter $\mu$. For display clarity, the curves have been vertical shifted. The measured $\mathcal{I}(\mu)\simeq 0.5011$ is independent of $\mu$ since the same random number are used to construct the lognormal cascade $\epsilon_{\mu}(t)$, and Brownian motion $B(t)$, for different $\mu$.}\label{fig:MRW}
\end{figure}

\subsubsection{Multifractal random walk with a discrete cascade}
Another important test case to detect the potential influence of the intermittency correction is MRW with lognormal statistics, which is defined as
\begin{equation}
x_{\mu}(t)=\int_0^t \epsilon_{\mu}(t')^{1/2}\upd B(t') , \label{eq:MRW}
\end{equation}
in which $\epsilon_{\mu}(t)$ is a multifractal measure with lognormal statistics to provide an intermittency correction $\mu$, and $B(t)$ is the Brownian motion to provide the scaling of the final process $x_{\mu}(t)$ \cite{Bacry2001,Muzy2002,Schmitt2003,Schmitt2016Book}. Figure~\ref{fig:MFmeasure} illustrates the cascade process algorithm.
The large scale corresponds to a unique cell of size $L=\ell_0 \lambda^p$, where $\ell_0$ is a fixed scale and $\lambda>1$ is the scale ratio, which for discrete models is typically set as $\lambda=2$. The next level subscale corresponds to $\lambda$ cells, each of which has the size $L/\lambda=\ell_0 \lambda^{p-1}$. Such scale cascade process continues from step $(1,2, ...)$ till $p$, leading to $\lambda^p$ cells in total with the size of $L/\lambda^p=\ell_0$, which is the smallest scale of the cascade. Finally, the multifractal measure is written as the following product of
$p$ cascade random variables,
 \red{\begin{equation}
  \epsilon_{\mu}(t)=\prod_{n=0}^p W_{n,t},
\end{equation}
where $W_{n,t}$ is the  lognormal random variable with independent identically distribution (i.i.d) corresponding to the position $t$ and level $n$ in the cascade with a mean value $\langle \ln W_{n,t} \rangle=-\frac{1}{2}\mu\ln \lambda$ and variance $\sigma^2( \ln W_n)=\mu\ln \lambda$, where $\mu$ characterizes the intermittency parameter \cite{Schmitt2003}}.

Figure \ref{fig:MRW} shows a synthesized MRW time series $x_{\mu}(t)$, with various intermittency parameter $\mu$ and the data length $L=2^{17}$ data points. Same as for fBm, in the synthesized algorithm the same random number was used to construct the lognormal cascade $\epsilon_{\mu}(t)$ and $B(t)$, respectively. Visually, all these curves have the similar trend with increasing relative variation with $\mu$. The corresponding experimental $\mathcal{I}(\mu)=1/2$ 
is independent with $\mu$.  A detail check suggests that  the location of local extremal points remains invariant for different $\mu$, showing that the intermittency correction does not change the distribution of extremal points, but change the amplitude.

\subsubsection{Structure-function scaling}

\begin{figure}[!htb]
\centering
 \includegraphics[width=0.85\linewidth]{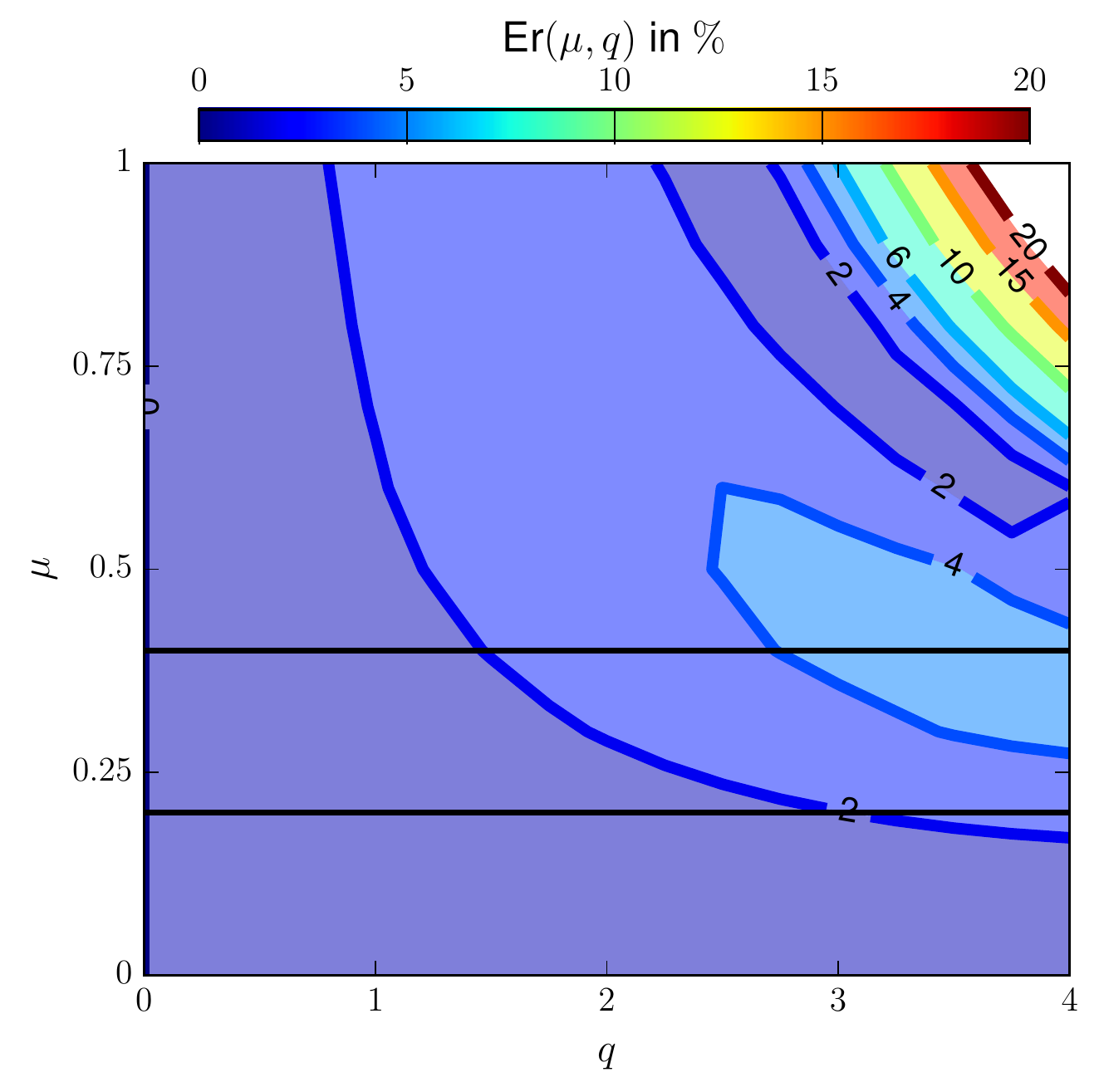}
  \caption{(Color online) Contour plot of the measured relative error between the structure function scaling exponent $\zeta(q)$ and the lognormal prediction Eq.\,\eqref{eq:lognormal}. A typical value $0.2\le \mu\le 0.4$ for the Eulerian turbulent velocity is indicated by a horizontal solid line. }\label{fig:MFerror}
\end{figure}

The important feature of the present MRW can be understood from the structure function scaling. Conventionally the $q$-th order structure function is defined as
\begin{equation}
S_q(\tau)=\langle \vert \Delta x_{\mu,\tau} (t)\vert^q\rangle\propto \tau^{\zeta(\mu,q)},
\end{equation}
where $\Delta x_{\mu,\tau}(t)=x_{\mu}(t+\tau)-x_{\mu}(t)$ is the increment, $\tau$ is the separation scale, and $\mu$ is the intermittency parameter ($0\le\mu\le1$) characterizing the lognormal cascade \cite{Schmitt2003}. With the lognormal $\epsilon_{\mu}(t)$,
the scaling exponent $\zeta(\mu,q)$ can be expressed as
\begin{equation}
\zeta(\mu,q)=\frac{q}{2}-\frac{\mu}{2}(\frac{q^2}{4}-\frac{q}{2}) , \label{eq:lognormal}
\end{equation}
which previously has been verified for the $\mu=0.15$ case by \citet{Huang2011PRE} and \citet{Huang2014JoT}.

The final $q$th-order structure function $S_q(\tau)$ results from an ensemble average over 100 MRW realizations with $p=17$, i.e. $131,072$ data points. The power-law behavior is observed for all different $\mu$ (not shown here), based on which the scaling exponent $\zeta(\mu,q)$ can be estimated in the range $100\le \tau \le 10,000$. Figure \ref{fig:MFerror} shows the contour plot of the relative error between the measured $\zeta(\mu,q)$ and the lognormal formula in Eq.\,\eqref{eq:lognormal}. Visually, except for the upper-right corner, the relative error is below $5\%$, verifying MRW as expected in a large range of parameter $\mu$. Note that for the classical Eulerian turbulent velocity, a typical intermittency parameter is found in the range $0.2\le \mu\le 0.4 $ \cite[see page 165]{Frisch1995}. Despite the Hurst number difference ($1/3$ for the Eulerian velocity, and $1/2$ for MRW here), the MRW model is heuristic to reproduce the same intermittency correction.

\subsubsection{Extremal-point-density statistics}
\begin{figure}[!htb]
\centering
 \includegraphics[width=0.85\linewidth]{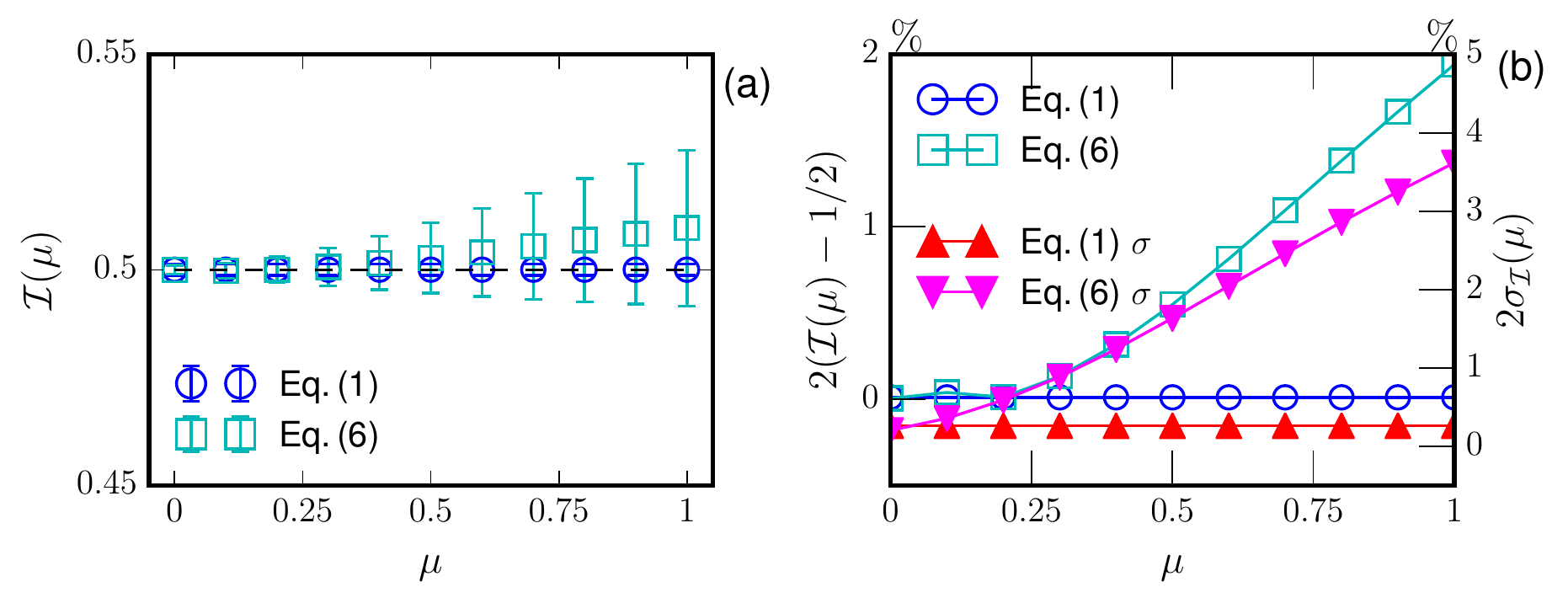}
  \caption{(Color online) a) Measured EPD $\mathcal{I}(\mu)$ for MRW. The errorbar is the standard deviation from 100 realizations. Due to the intermittency effect, Eq.\,\eqref{eq:TEPD} overestimates $\mathcal{I}(\mu)$. b) The  relative error $\mathrm{Er}(\mu)$ \red{for the measured $\mathcal{I}(\mu)$ and the standard deviation $\sigma$.}.}\label{fig:MFEP}
\end{figure}

\begin{figure}[!htb]
\centering
 \includegraphics[width=0.85\linewidth]{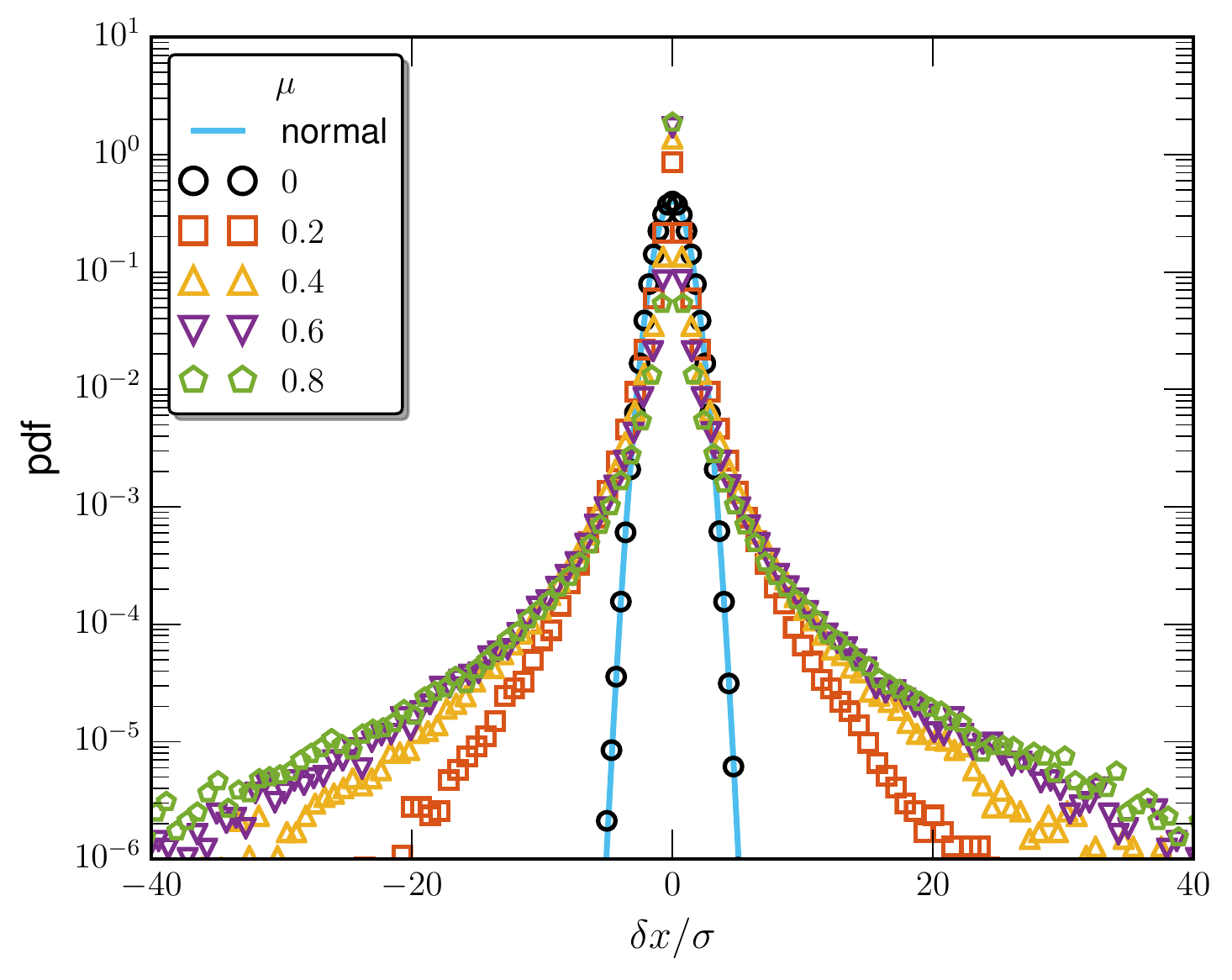}
  \caption{(Color online) Experimental pdfs for the first-order derivative of MRW, $\delta x=x'_{\mu}(t)$, with various $\mu$. For display clarity, the pdfs have been normalized by their respective standard deviation.}\label{fig:MpdfIncrement}
\end{figure}

Figure \ref{fig:MFEP}\,(a) shows $\mathcal{I}(\mu)$ from direct counting ($\ocircle$) and Eq.\,\eqref{eq:TEPD} ($\square$), \red{(b) the measured relative error for $\mathcal{I}(\mu)$ and the standard deviation obtained from 100 realizations, which is shown as errorbar in (a).} There is an overestimation from Eq.\,\eqref{eq:TEPD}. The relative error $\mathrm{Er}(\mu)$ increases with $\mu$ from $0\%$ up to $2\%$, and the corresponding standard deviation increases from $0\%$ up to $4\%$, which can be ascribed to the violation of the joint Gaussian distribution. As demonstrated in Fig.\,\ref{fig:MpdfIncrement}, except for the Brownian motion case with $\mu=0$, the measured pdfs of $x'_{\mu}(t)$, the first-order derivative of $x_{\mu}(t)$, are far from the normal distribution. Moreover, the tail part of the pdf increases with $\mu$, implying a stronger intermittency when $\mu$ is larger.

\subsubsection{Coarse-grained effect}

\begin{figure}[!htb]
\centering
 \includegraphics[width=0.85\linewidth]{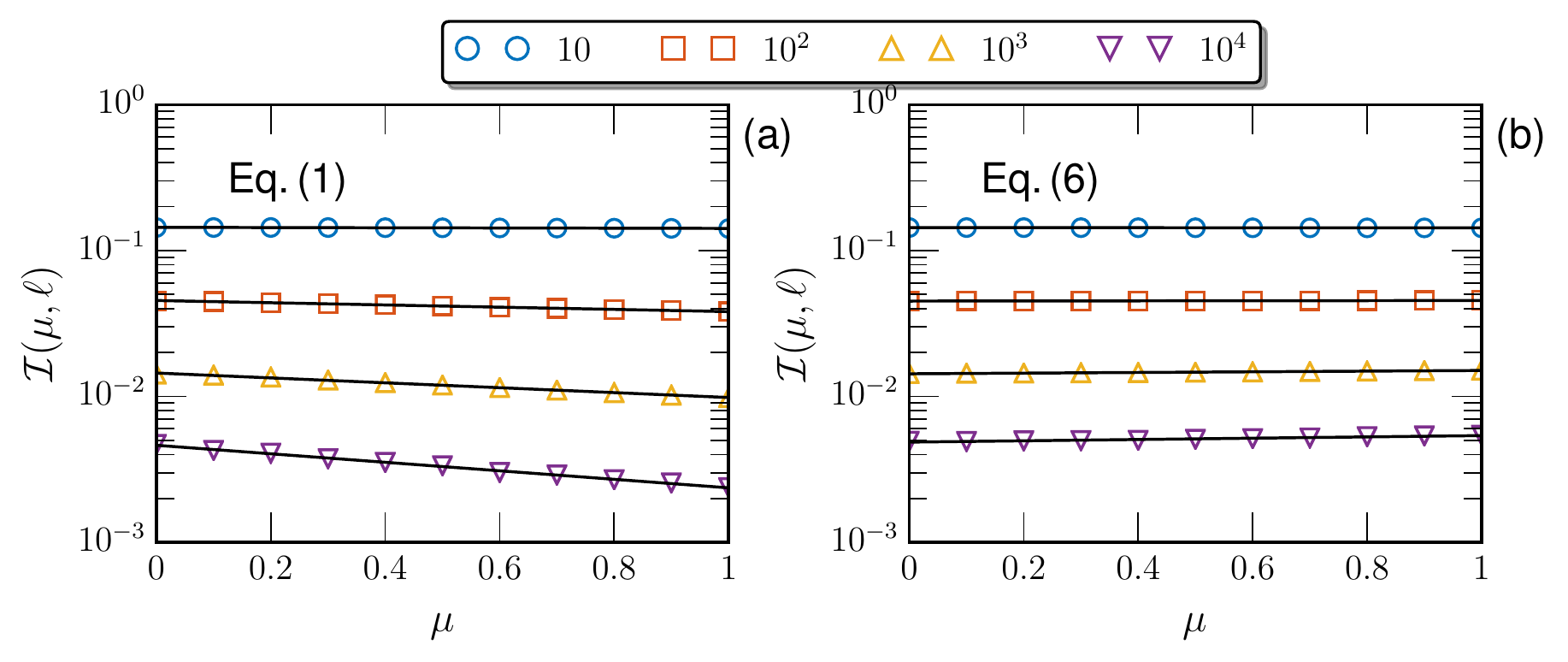}
  \caption{(Color online)  Measured $\mathcal{I}(\mu,\ell)$ versus $\mu$ at various scale $\ell$ from a) direct counting by Eq.\,\eqref{eq:extrema}, b) Eq.\,\eqref{eq:TEPD}. The solid line is an exponential law fitting. Exponential-law is observed for both approaches, but with different trends.  }\label{fig:MFEPDCG}
\end{figure}

\begin{figure}[!htb]
\centering
 \includegraphics[width=0.85\linewidth]{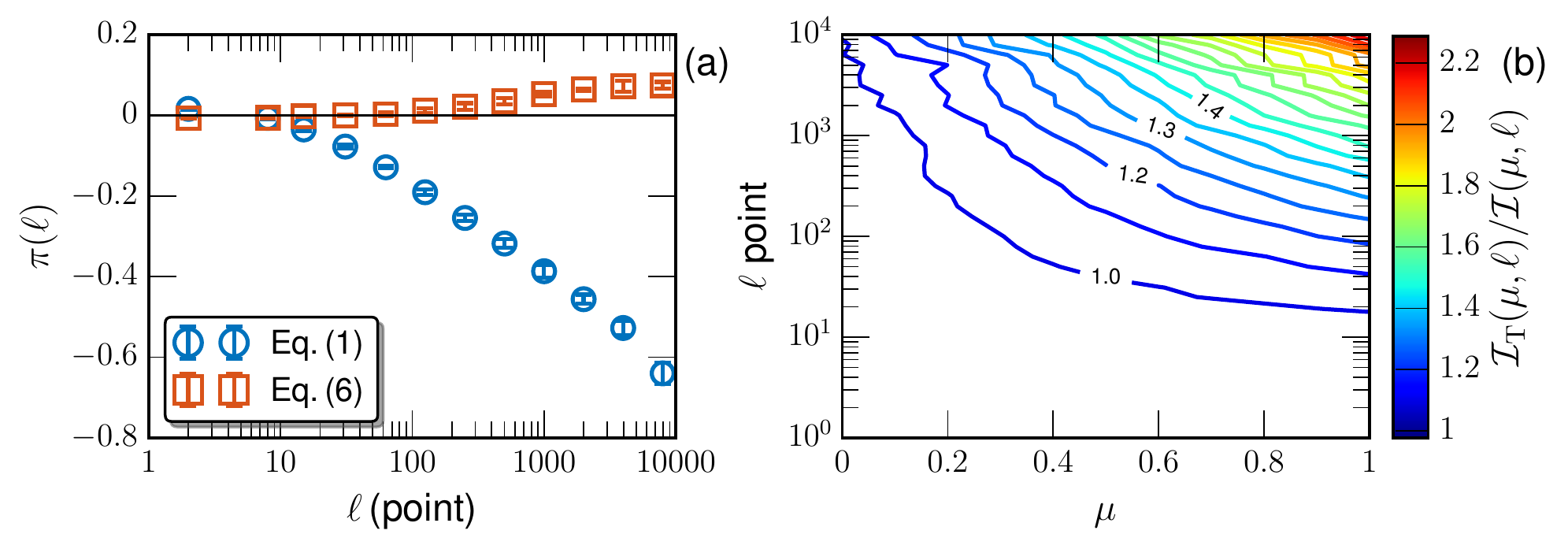}
  \caption{(Color online) a) Experimental scaling exponent $\pi(\ell)$ versus $\ell$. Due to the violation of the joint Gaussian distribution requirement, Eq.\,\eqref{eq:TEPD} fails to measure $\pi(\ell)$. b) Contour plot of the measured ratio $\mathcal{I}_{\mathrm{T}}(\mu,\ell)/\mathcal{I}(\mu,\ell)$ to confirm the overestimation of $\mathcal{I}(\mu,\ell)$ by Eq.\,\eqref{eq:TEPD}. }\label{fig:MFEPDCG2}
\end{figure}

\begin{figure}[!htb]
\centering
 \includegraphics[width=0.85\linewidth]{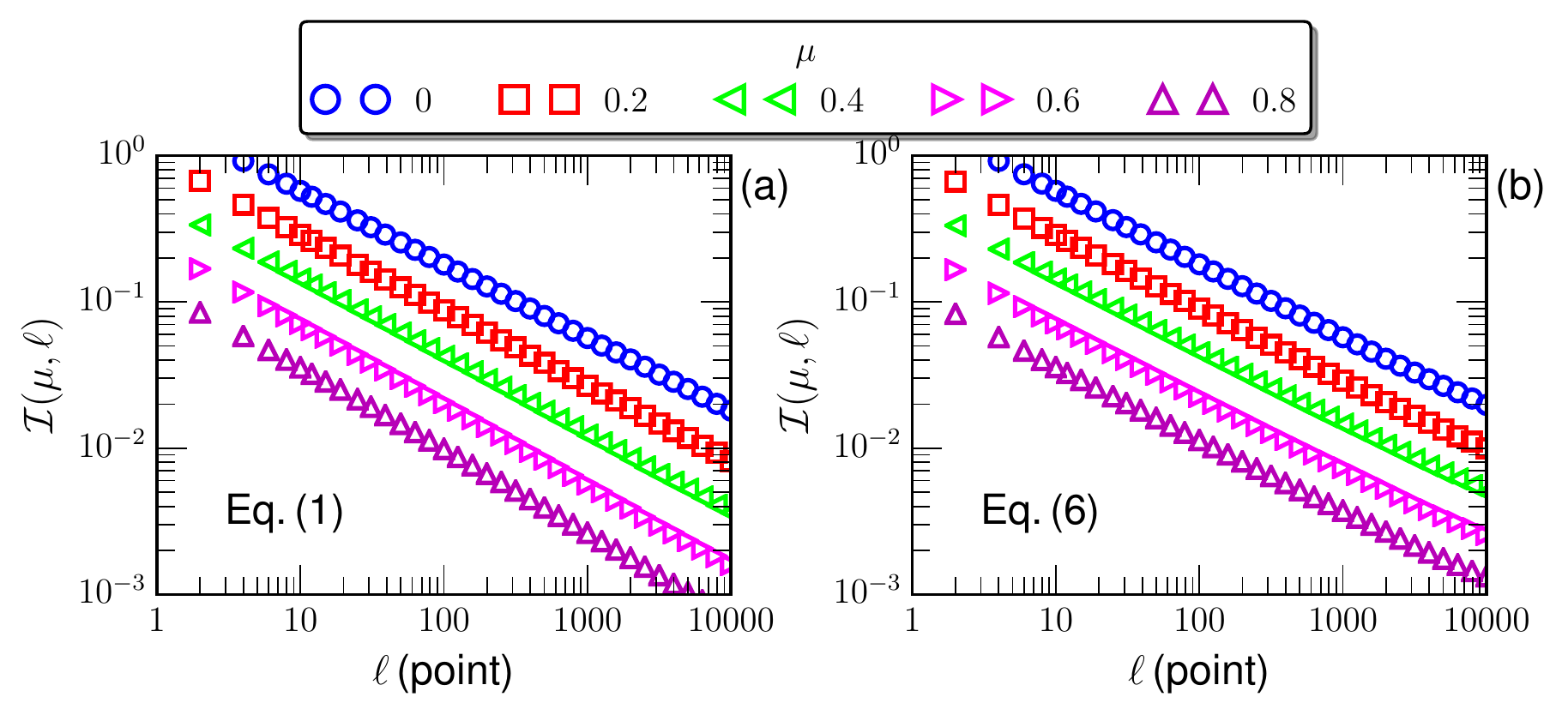}
  \caption{(Color online) $\mathcal{I}(\mu,\ell)$ versus $\ell$ from a) direct counting, and b) Eq.\,\eqref{eq:TEPD}. For display clarity, the curves have been vertical shifted. The corresponding scaling exponent $\xi(\mu)$ is fitted in the range $100\le\ell\le 1,000$.}\label{fig:MFEPScaling}
\end{figure}

\begin{figure}[!htb]
\centering
 \includegraphics[width=0.85\linewidth]{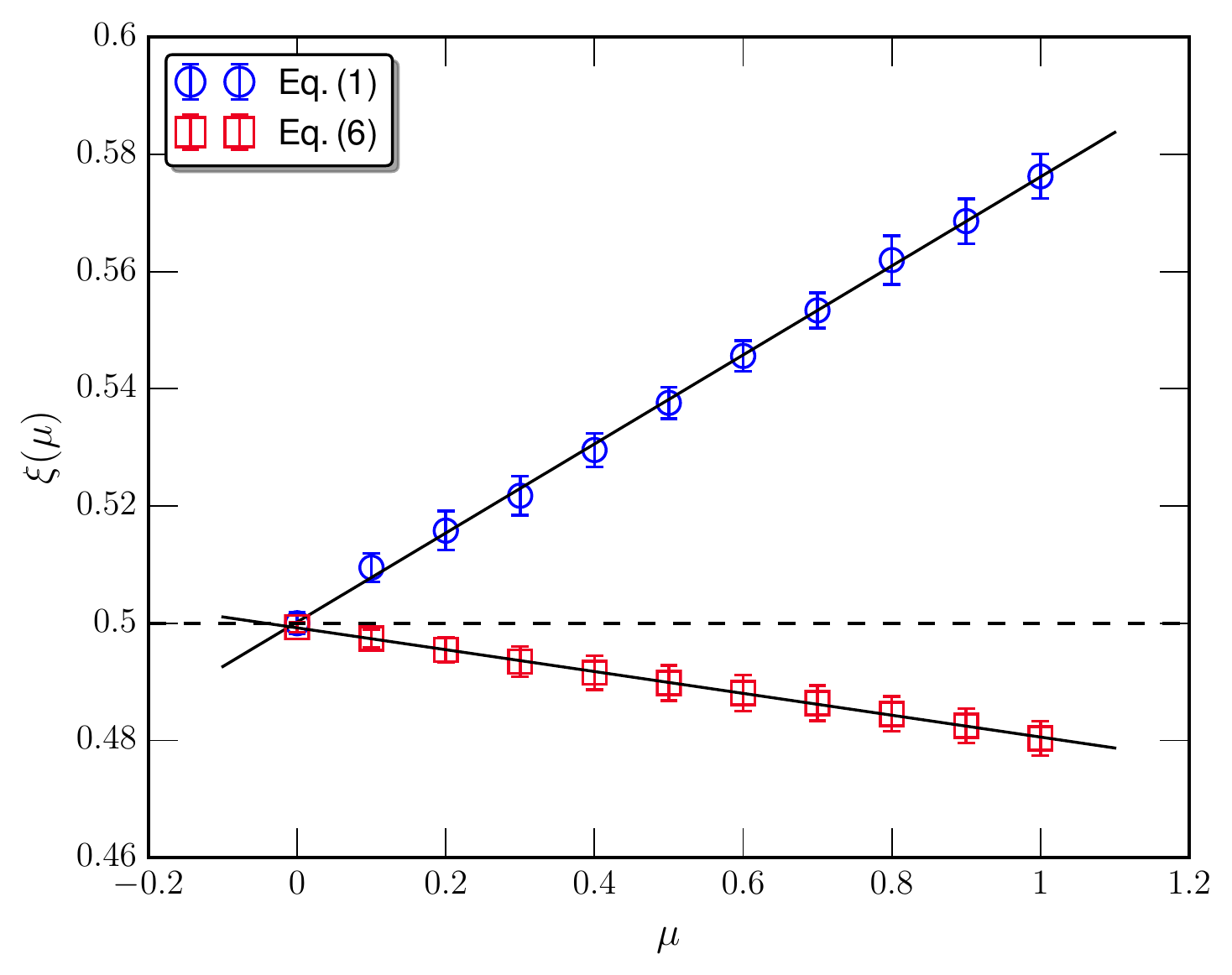}
  \caption{(Color online) Scaling exponent $\xi(\mu)$ versus $\mu$ from a) direct counting ($\ocircle$), and b) Eq.\,\eqref{eq:TEPD} ($\square$), where a linear fitting is illustrated by a solid line. The horizontal dashed line indicates the value $1/2$ for Brownian motion.}\label{fig:MFEPScaling2}
\end{figure}

Figure \ref{fig:MFEPDCG} shows $\mathcal{I}(\mu,\ell)$ versus $\mu$ at several $\ell$ from a) direct counting, and b) Eq.\,\eqref{eq:TEPD}. The following exponential-law behavior is observed
\begin{equation}
\mathcal{I}(\mu, \ell)\propto e^{\pi(\ell)\mu},
\end{equation}
where $\pi(\ell)$ is a $\ell$-dependent scaling exponent. Visually, two approaches provide opposite trends. More precisely, the result from direct counting predicts an exponential decay with $\mu$, while the second estimator provides an exponential growing.

Figure \ref{fig:MFEPDCG2}\,a) shows the experimental $\pi(\ell)$ provided by direct counting (Eq.\,\eqref{eq:extrema}, $\ocircle$) and Eq.\eqref{eq:TEPD} ($\square$), confirming the observation in Fig.\,\ref{fig:MFEPDCG}. Moreover, the intensity (absolute value) of scaling exponents $\pi(\ell)$ provided by Eq.\,\eqref{eq:TEPD} is much smaller than the one by direct counting for large values $\ell$. Figure \ref{fig:MFEPDCG2}\,b) shows a contour plot of the measured ratio $\mathcal{I}_{\mathrm{T}}(\mu,\ell)/\mathcal{I}(\mu,\ell)$, showing that Eq.\,\eqref{eq:TEPD} overestimates $\mathcal{I}(\mu,\ell)$ more when $\ell$ and $\mu$ increase.

Figure \ref{fig:MFEPScaling} shows the measured EPD for various $\mu$ from a) direct counting and b) Eq.\eqref{eq:TEPD}, where for display clarity the curve has been vertical shifted. The power-law behavior is observed for all $\mu$. The scaling exponent $\xi(\mu)$ is estimated in the range $100\le \ell\le 1,000$.

Figure \ref{fig:MFEPScaling2}  shows the measured $\xi(\mu)$ versus the intermittency parameter, $\mu$. Using direct counting the measured $\xi(\mu)$ increases linearly with $\mu$ with an experimental slope $\simeq 0.08$, while due to the violation of the joint Gaussian distribution requirement, the measured $\xi_{\mathrm{T}}(\mu)$ by Eq.\,\eqref{eq:TEPD} decreases linearly with $\mu$ with a slope $\simeq-0.02$.

According to the obtained results, it is meaningful to comment on the scaling exponent $\xi$. For a mono-fractal process, the measured $\xi$ is found to be the same as the Hurst number; or in other words, Eq.\,\eqref{eq:PL} provides a new idea to estimate the Hurst number.  For a scaling process
with intermittency correction, as shown in this work, $\xi$ could also be influenced by the process intermittency, where Eq.\,\eqref{eq:TEPD} overestimates $\mathcal{I}_{\mathrm{T}}$, but underestimates $\xi$.

\section{Applications in Turbulent systems}\label{sec:Applications}

\begin{figure}[!htb]
\centering
 \includegraphics[width=0.85\linewidth]{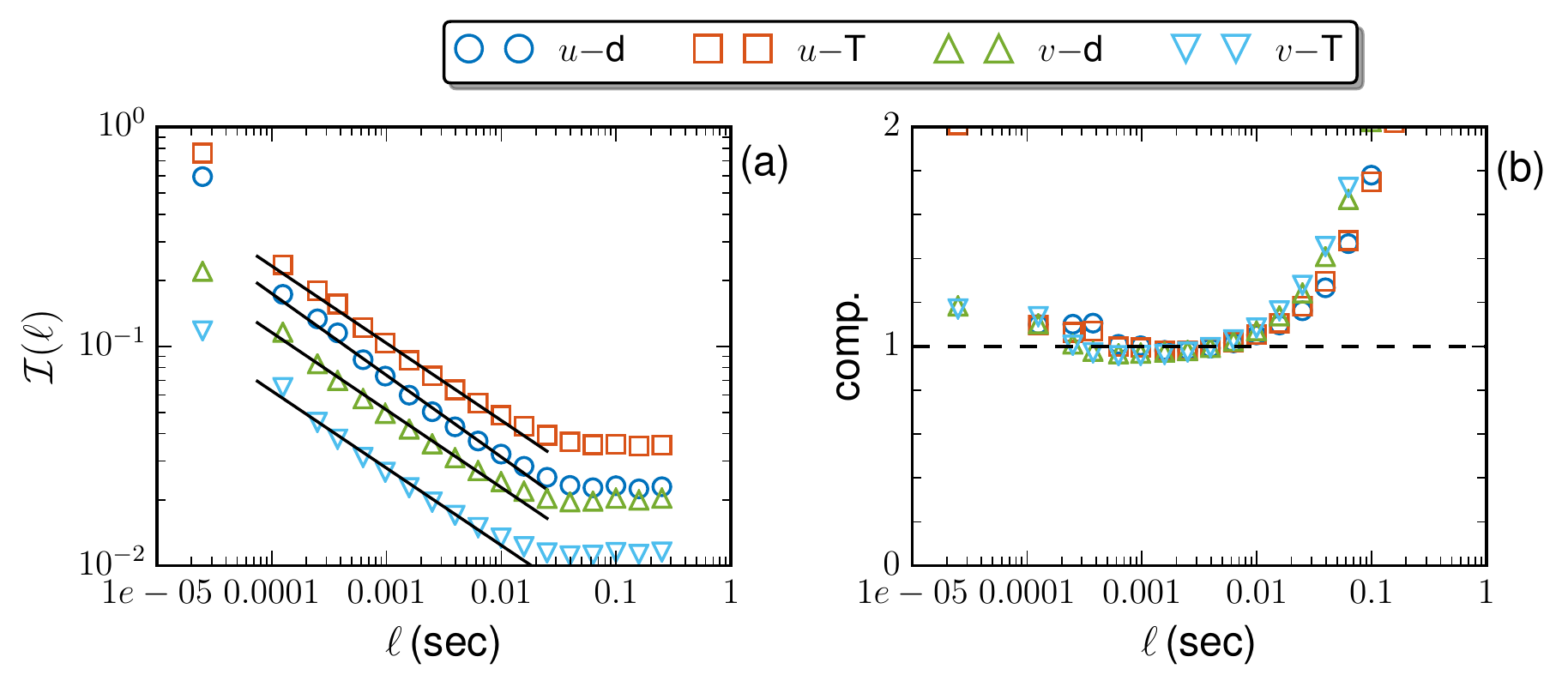}
  \caption{(Color online) a) Measured  $\mathcal{I}_H(\ell)$ for the turbulent velocity obtained from a wind tunnel experiment \cite{kang2003} with a Reynolds number $Re_{\lambda}\simeq 720$. For display clarity, the curve for $u-$T and $v-$T have been vertically shifted by multiplying $1.5$ and $0.5$, respectively. The solid line is the power-law fit in the range $0.0001\le \tau \le 0.01\,$sec, corresponding to a frequency in the range $100<f<10,000\,$Hz. b) The corresponding curves compensated by the fitting parameters. }\label{fig:JHUEPD}
\end{figure}

\subsection{Eulerian turbulent velocity in wind tunnel}

First we consider here a velocity database obtained from a wind tunnel experiment with Taylor scale $\lambda$ based Reynolds number as high as $Re_{\lambda}\simeq720$~\cite{kang2003}. A probe array with four X-type hot wire anemometry were placed in the middle height and along the center line of the wind tunnel to record the velocity with a sampling frequency of $40$\,kHz at the streamwise direction $x/M=20$, in which $M$ is the size of the active grid. The measure time is 30 seconds with 30 times repetition, i.e. totally there are $30\times 4\times 30\times (40\times 10^3)$ data points. The Fourier power spectrum $E_u(k)$ of the longitudinal velocity reveals a nearly two decades inertial range in the frequency range $10\le f\le1,000\,$Hz (i.e. the time scale $0.001<\tau<0.1\,$sec) with a scaling exponent $\beta\simeq 1.65\pm0.02$. More details about this database can be found in Ref.\,\cite{kang2003}.
Let us mention that the Taylor\rq{}s frozen hypothesis~\cite{Frisch1995} is not implemented here to convert the results into the spatial coordinate.

Figure \ref{fig:JHUEPD}\,a) shows the measured EPD for both longitudinal $u$ and transverse $v$ velocity components via direct counting (denoted as d) and Eq.\,\eqref{eq:TEPD} (denoted as T). For display convenience, the curves have been vertically shifted. The clear power-law behavior exists in the range $0.0001\le\tau\le0.01\,$sec, corresponding to a frequency range $100\le f\le 10,000\,$Hz. Note that the scaling range here is different from the one predicted by the Fourier power spectrum since the coarse-grained operator has been applied to measure $\mathcal{I}(\ell)$ \citep{Kraichnan1974JFM}. It is found that $\xi=0.37\pm0.01$ for the $u$-d and $0.35\pm0.01$ for the rest. Interestingly for the same database the $0.37$ scaling has been observed from the velocity increment pdf \citep{Huang2011PoF}, which also agrees with the first-order structure function scaling for high Reynolds number turbulent flows \citep{Benzi1993PRE}.

\subsection{Eulerian velocity in turbulent boundary layer}
\begin{figure}[!htb]
\centering
 \includegraphics[width=0.85\linewidth]{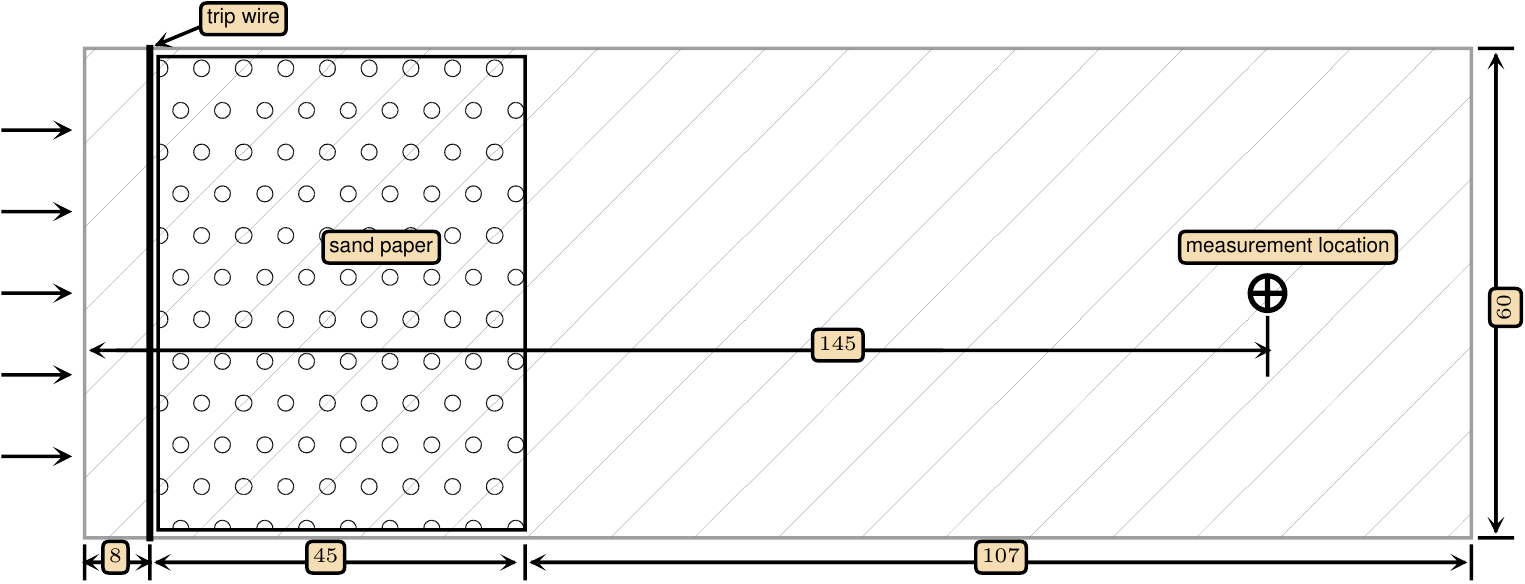}
  \caption{(Color online) Schematic of the  boundary layer experimental setup ~\citep{Xiao2015CPB} with a unit cm. The inflow is $4.5\,\mathrm{ms}^{-1}$ with a Reynolds number $Re_{\theta}\simeq810$. A trip-wire with a twisted-wire is located at $8\,$cm downstream to accelerate the turbulent boundary layer development.  The measurement is performed at $145\,$cm downstream ($\oplus$). }\label{fig:TJUES}
\end{figure}

\begin{figure}[!htb]
\centering
 \includegraphics[width=0.85\linewidth]{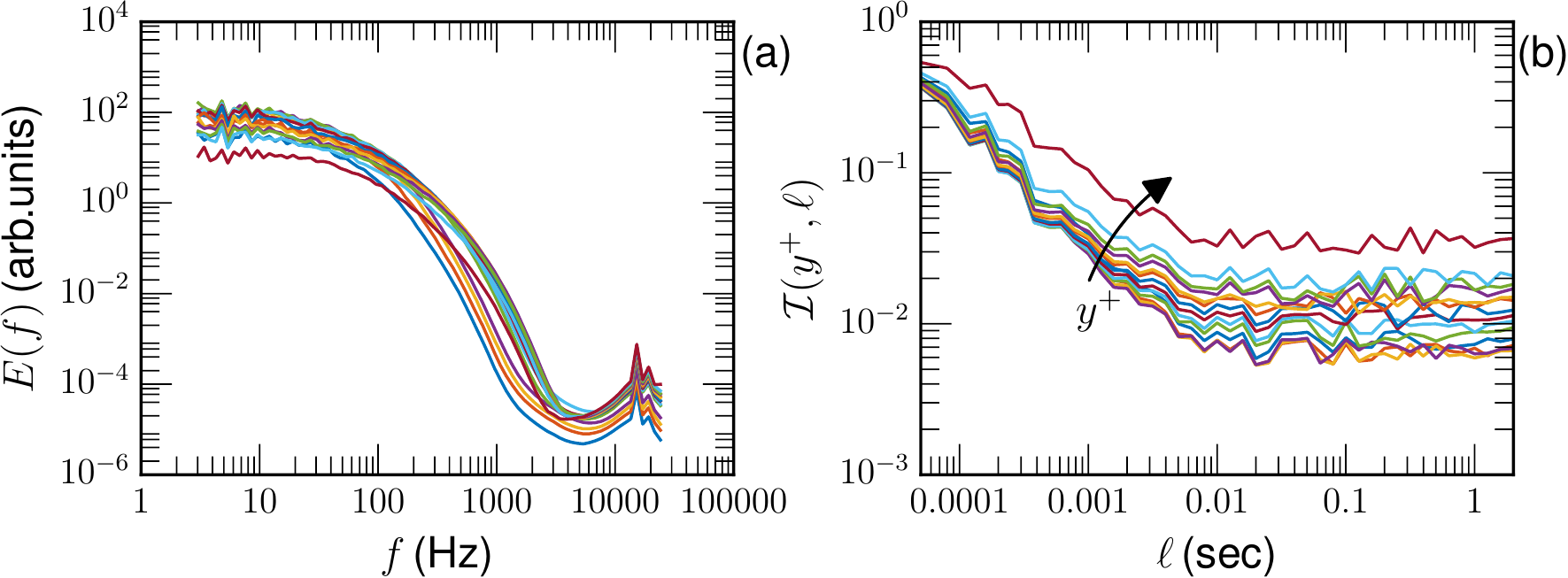}
  \caption{(Color online)  a) Measured Fourier power spectrum at different height $y^+$. b) The corresponding $\mathcal{I}(y^+,\ell)$ versus $\tau$.  }\label{fig:TJUFPS}
\end{figure}

\begin{figure}[!htb]
\centering
 \includegraphics[width=0.85\linewidth]{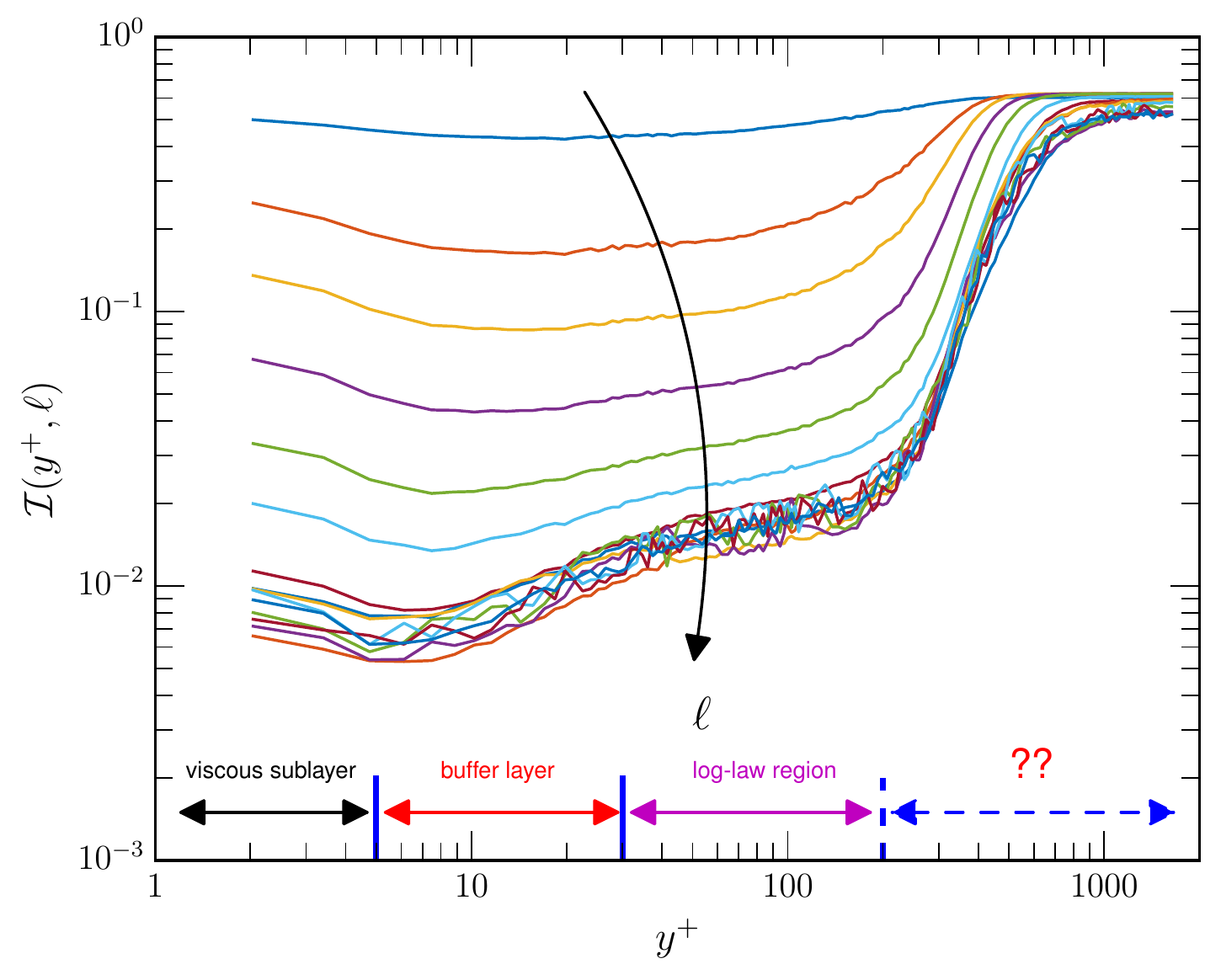}
  \caption{(Color online) The measured EPD $\mathcal{I}(y^+,\tau)$ versus $y^+$ reproduced in a log-log view, together with the four regimes of the turbulent boundary layer structure.}\label{fig:TJUEPD}
\end{figure}

Another turbulent data to be analyzed is the Eulerian velocity from a zero-pressure-gradient turbulent boundary layer experiment~\citep{Xiao2015CPB}. We recall briefly the main parameters involved. As shown in the schematic of experiment setup in figure \ref{fig:TJUES}, to achieve a fully developed boundary layer structure, a trip wire with diameter $2\,\mathrm{mm}$ is placed at $8\,\mathrm{cm}$ after the leading edge, followed with a $45\,\mathrm{cm}$ in length sand paper. The inflow speed is $4.5\,\mathrm{ms}^{-1}$ and the measurement is performed at $145\,\mathrm{cm}$ downstream. The corresponding momentum thickness $\theta$ based  Reynolds number is $Re_{\theta}\simeq 810$.
A commercial  hot-wire is operated in constant-temperature anemometry mode by TSI-IFA300 unit. The signals are sampled in $\sim84\,$sec at frequency  $f_s=50\,\mathrm{kHz}$ with a low-pass at a frequency of $25\,\mathrm{kHz}$. We consider only the longitudinal velocity.

Figure \ref{fig:TJUFPS}\,a) shows the measured Fourier power spectrum $E(f)$ at different height $y^+$, where the cutoff frequency is roughly around $f_N=3,000\,$Hz, above which the data is dominated by noises. Due to the finite Reynolds number, no clear power-law behavior can be observed. Figure \ref{fig:TJUFPS}\,b) shows the measured $\mathcal{I}(y^+,\ell)$ versus $\ell$. In contrast the power-law can be observed roughly when $\ell\le 0.01\,$sec. However, such difference could be due to the measurement noise. The measured $\mathcal{I}(y^+,\ell)$ decreases rapidly with $\ell$ and becomes saturated when $\ell\gg0.01\,$sec. For a fixed $\ell$, $\mathcal{I}(y^+,\ell)$ seems to increase with $y^+$.

We then reproduce the measured $\mathcal{I}(y^+,\ell)$ versus $y^+$ in Fig.\,\ref{fig:TJUEPD}. Different regimes of the boundary layer are indicated by vertical lines. They are viscous sublayer with $y^+\le 5$, buffer layer with $5\le y^+\le 30$, log-law region with $30\le y^+\le 200$, and outer layer $y^+>200$, respectively. Interestingly the measured EPD for large value of $\ell$ shows four different regimes as well, coincidentally agreeing with the four different boundary layer regimes. Therefore it is reasonable to claim that the proposed EPD analysis can be effective to detect the boundary layer structure. Additionally the inverse of EPD,  $\mathcal{I}^{-1}(y^+,\ell)$, roughly measures an average scale of turbulent structure. Therefore, a small value of $\mathcal{I}(y^+,\ell)$ indicates some well-organized large-scale structure.

\section{Conclusions}\label{sec:Conclusion}
In summary, EPD of several typical scaling time series has been investigated. For fractional Brownian motion case, the result  agrees with the the theoretical prediction by \citet{Toroczkai2000PRE} (Eqs.\,\eqref{eq:TEPD} \red{or} \eqref{eq:MF1}) since the process satisfies the Gaussian condition of the joint distribution of the correlation coefficients. 
When the Hurst number $H\le 0.6$, the measured EPD agrees with the formula\,\eqref{eq:MF3}, but deviates when $H>0.6$. Using a coarse-grained operator the measured scaling exponent $\xi(H)$ is found to equal to $H$, providing a new idea to estimate the Hurst number.
\red{Due to the non-differentiable property of the fBm process, EPD predicted by the Rice's formula is largely different from the direct counting result. } 
 For multifractal random walk with lognormal statistics, the EPD via direct counting is independent with the intermittency parameter $\mu$, suggesting that the intermittency effect may not change the distribution of the extremal points, but change the amplitude. Due to the intermittency correction, Eq.\,\eqref{eq:TEPD} overestimates EPD systematically, which is due to the violation of the joint Gaussian distribution requirement. After coarse-grained operation, the power-law behavior is still preserved. The result from direct counting suggests that $\xi(\mu)$ increases linearly with $\mu$; differently $\xi(\mu)$ from Eq.\,\eqref{eq:TEPD} decreases linearly with $\mu$.

The EPD analysis was then applied to the experimental turbulent velocity data from a high Reynolds number wind tunnel flow with $Re_{\lambda}\simeq 720$ and a turbulent boundary layer with $Re_{\theta}\simeq 810$. For the former case, the scaling exponent $\xi$ for the longitudinal velocity is $\xi\simeq 0.37$, which is in agreement with the value from the conventional first-order structure function analysis via the extended self-similarity technique \cite{Benzi1993PRE}. For the latter one, the measured EPD after the coarse-grained operation shows clearly four regimes, which coincides  with the classical turbulent boundary layer structure, including the viscous sublayer, buffer layer, log-law region and the outer layer. A high-order dimension (resp. 2D and 3D) extension of this approach for PIV (particle image velocimetry) measurement or high resolution DNS (direct numerical simulation) is under progress, and will be shown elsewhere.

\begin{acknowledgments}
This work is sponsored by the National Natural Science Foundation of China (under Grant Nos. 11332006 and 91441116), and  partially by the Sino-French (NSFC-CNRS) joint research project (No. 11611130099, NSFC China, and PRC 2016-2018 LATUMAR ``Turbulence lagrangienne: \'etudes num\'eriques et applications environnementales marines",  CNRS, France).  Y.H. is also supported by the Fundamental Research Funds for the Central Universities (Grant No. 20720150075).
We thank Prof. Meneveau at Johns Hopkins University  for  providing us the experiment data, which can be found at: \footnote{\href{http://turbulence.pha.jhu.edu/}{http://turbulence.pha.jhu.edu/}}.
A \textsc{Matlab} source package to realize the Extremal-Point-Density  analysis  is available at: \footnote {\href{https://github.com/lanlankai}{https://github.com/lanlankai}}. \red{Useful comments by one referee are gratefully acknowledged.}
\end{acknowledgments}

%

\end{document}